\documentclass[fleqn,usenatbib]{mnras}

\usepackage{newtxtext,newtxmath}

\usepackage[T1]{fontenc}

\DeclareRobustCommand{\VAN}[3]{#2}
\let\VANthebibliography\thebibliography
\def\thebibliography{\DeclareRobustCommand{\VAN}[3]{##3}\VANthebibliography}

\usepackage{graphicx}	
\usepackage{amsmath}	
\usepackage{array}      
\usepackage{tabularx}   

\title[Mutual coupling in the MWA]{Investigating mutual coupling in the MWA Phase II compact array}

\author[K. Elder et al.]{
Katherine Elder,$^{1}$\thanks{E-mail: kat.elder@asu.edu}
Daniel C. Jacobs,$^{1}$
Maria Kovaleva$^{2}$
\\
$^{1}$School of Earth and Space Exploration, Arizona State University, Tempe, AZ, USA\\
$^{2}$International Centre for Radio Astronomy Research, Curtin University, Bentley, WA 6102, Australia
}

\date{Accepted XXX. Received YYY; in original form ZZZ}

\pubyear{\the\year{}}

\begin{document}
\label{firstpage}
\pagerange{\pageref{firstpage}--\pageref{lastpage}}
\maketitle

\begin{abstract}
Measurement of the power spectrum of high redshift 21 cm emission from neutral hydrogen probes the formation of the first luminous objects and the ionization of intergalactic medium by the first stars. However, the 21 cm signal at these redshifts is orders of magnitude fainter than astrophysical foregrounds, making it challenging to measure. Power spectrum techniques may be able to avoid these foregrounds by extracting foreground-free Fourier modes, but this is exacerbated by instrumental systematics that can add spectral structure to the data, leaking foreground power to the foreground-free Fourier modes. It is therefore imperative that any instrumental systematic effects are properly understood and mitigated. One such systematic occurs when neighboring antennas have undesired coupling. A systematic in Phase II data from the MWA was identified which manifests as excess correlation in the visibilities. One possible explanation for such an effect is mutual coupling between antennas. A numerical electromagnetic software simulation of the antenna beam using FEKO has been built to estimate the amplitude of this effect for multiple antennas in the MWA. The numerical model predicts an amplitude which exceeds the requirement to avoid spreading the foreground. More work is necessary to better validate the required level of coupling and to verify that approximations did not under estimate the level of coupling. 
\end{abstract}

\begin{keywords}
dark ages, reionization, first stars -- instrumentation: interferometers -- methods: analytical
\end{keywords}



\section{Introduction}

The redshifted 21 cm emission from neutral hydrogen signal is a key probe of the Cosmic Dawn and Epoch of Reionization (EoR), the period of cosmic history when the first stars and galaxies formed and reionized the intergalactic medium. Detecting this signal will provide insight into the evolution of stars and galaxies. There are a number of radio interferometry experiments currently targeting the cosmological 21 cm power spectrum, including the Murchison Widefield Array \citep[MWA;][]{MWATingay2013, MWAPhase2}, the Hydrogen Epoch of Reionization Array \citep[HERA;][]{HERA2017, Berkhout2024}, and the Square Kilometre Array-Low \citep[SKA-Low;][]{SKALow}. 

One of the greatest challenges in detecting the 21 cm signal is distinguishing it from astrophysical foregrounds, which are 10 to $10^8$ times brighter than the 21 cm signal \citep[][]{Santos2005}. Because foreground emission is spectrally smooth while 21 cm emission is not, foregrounds are largely isolated to line of sight power spectrum modes mapping to delays shorter than the baseline length. This ``wedge'' in the 2D power spectrum leaves a region where the 21 cm signal could be most easily observed \citep[][]{Trott2012,Parsons2012,Liu2014a,Liu2014b,Thyagarajan2015,Chapman2016}. Further subtraction of foreground power can potentially open more power spectrum modes. Subtraction of a known sky model must account for the telescope's response \citep[][]{BonaldiBrown,Chapman2016,Pober2016}. Systematic effects which introduce unexpected spectral structure can cause the foregrounds to contaminate the EoR window and obscure the 21 cm signal. Precise understanding of the telescope's beam and analog system is required. It is estimated that systematics must be reduced to approximately -70 dB below the 21 cm signal in visibility units in order to avoid this leakage \citep[][]{Trott2020}. 

Instrumental features can cause smooth spectrum foregrounds to appear to have spectral structure and simulation methods have been developed to study the impacts of antenna design, malfunctions, and other uncertainties \citep[][]{Cumner2023, Chokshi2024}. \citet{OHara2024} provides a timely and comprehensive review of work in this area to ask the question "How accurately, and in which regions of the observational domain, must one model the antenna system?" The conclusion of the literature study is that mutual coupling is likely a serious limiting factor. A detailed simulation provides a quantitative estimate of the systematic power spectrum bias that might be seen by SKA-low given the predicted levels of mutual coupling within each station. Coupling between stations is not included, nor is a comparison to data yet available from the newly built SKA stations.  

Meanwhile, operating arrays might provide some hints.
\citet{Kolopanis2023} reported an unidentified systematic while investigating the shortest baselines in the MWA Phase II compact array. These baselines had previously been ignored because they did not calibrate well, however short baselines are the most sensitive to the EoR signal, therefore further investigation is warranted. One of the hypothetical causes of this systematic is mutual coupling between tiles. This paper examines this hypothesis. 

The MWA, operating in the frequency range 80-300 MHz, is an interferometer with phased array tiles of bow-tie dipoles as the primary antennas. Given that the electrical length of the shortest separation between the tiles is $(4-14)\lambda$, thus far, it has been thought for the tiles to have relatively low, and therefore negligible, levels of mutual coupling. However, a quantitative analysis has not previously been shown in the literature.

In radio astronomy, mutual coupling between antennas has recently received more attention as a potential source of error in 21 cm instruments. Current upper limits for 21 cm experiments report excess correlation above the expected noise level, and many include mutual coupling as a potential contributing factor left for further study \citep[][]{PAPER_limit, MWA_limit_2016, Kern2019, Kern2020, LOFAR_limits}. In discussion of the origin of systematic biases, MWA results do not point to mutual coupling as a potential source of systematic \citep[][]{Beardsley2015, Trott2020}. Informally it has been proposed by the MWA group that mutual coupling between MWA tiles has never been detected and is not a relevant problem for this instrument. The MWA does have several novel points which might minimize coupling. Bow-tie antennas present a low profile with much less than a wavelength above the ground. Tiles are laid out at spacings of  several wavelengths and phased arrays should provide some self-shielding compared to a single antenna receiver.  However until now the matter has not been studied. Here we address the knowledge gap by estimating the level of mutual coupling between MWA tiles. This estimate can then be compared against an estimated order of magnitude threshold requirement established based on predicted signal levels. If the coupling is high enough, further detailed study will be warranted.

Models of coupling have been explored for HERA \citep[][]{Kern2019, Kern2020, Rath2024, Fagnoni2021}, HIRAX \citep[][]{Sampath2024}, and the SKA \citep[][]{Kyriakou2025, Borg2020, OHara2024}. All adopt approximations necessary to make the problem tractable and in some cases estimate the resulting power spectrum bias. For example, \citet{OHara2024} investigate the effects of coupling within an SKA but not between stations. This allows the use of existing visibility simulators. In contrast, in this paper we simulate coupling between stations, estimating coupling from S-parameters, but stop short of an analysis that generates simulated visibilities and puts the results through a realistic data analysis pipeline to measure the resulting power spectrum bias. The use of simulations in predicting the instrument response is complicated by the difficulty in generating visibility simulations at sufficient accuracy; verifying that observed power spectrum excesses are not the result of numerical or software errors requires extensive and laborious validation \citep[][]{Aguirre2022,Kittiwisit2023, Line2025}. Here we ask whether further effort in this area is justified for the MWA.

Electromagnetic propagation within an array of antennas can be difficult to model analytically, making numerical electromagnetic (EM) simulations the gold standard for obtaining an accurate model \citep[][]{BuiVan2017,BuiVan2018, OHara2024, Sampath2024, Gueuning2024}. However, the computational demands of such simulations for a large array are high. FEKO running on a high-end desktop with 11 parallel CPUs and 128GB of RAM is able to calculate S-parameters for three tiles, for a total of 48 dipoles, at 221 frequency points in $\sim$19 hours. Larger models require the use of more powerful servers and even High Performance Computing services. This limits the amount of parameter space that can be explored, motivating an approximate method and simplified models. 

A recently developed solver, Fast Array Simulation Tool (FAST) \citep[][]{Gueuning2024}, has shown to be comparable to other EM solvers while taking only minutes rather than days or weeks for compact arrays. This tool may prove useful for future studies of mutual coupling in the MWA once it is made public. 



In this paper we are asking the question: "what is the level of mutual coupling between MWA tiles?" This question can be broken into component parts which have not, to our knowledge, yet been answered in the literature. What are the coupling coefficients between individual dipoles and the net coupling between tiles? Are there aspects to consider when evaluating coupling between separated phased array stations?  Are the MWA tiles particularly insensitive to mutual coupling? Here we make a first estimate of the coupling level and compare with an order-of-magnitude calculation for how much coupling can be tolerated. Using a set of smaller simulations we test for simplified physical models and then test for a large scale effects with a 21 tile simulation. 

This paper is structured as follows: Section~\ref{sec:MWA} provides an overview of the MWA Phase II; Section~\ref{sec:coupling} introduces mutual coupling and the various sources of coupling in interferometers; Section~\ref{sec:model} provides a detailed description of the electromagnetic model and the computational limits for this analysis; Section~\ref{sec:calc_vari} presents the results from the simulations; Section~\ref{sec:SvB} analyzes S-parameters as a function of baseline;
Section~\ref{sec:discussion} addresses limitations and open questions; and Section~\ref{sec:conclusion} summarizes the results.

\section{Murchison Widefield Array}\label{sec:MWA}

 The Murchison Widefield Array (MWA) is a radio interferometer located at Inyarrimanha Ilgari Bundara, the CSIRO Murchison Radio-astronomy Observatory, 800 kilometres (500 mi) north of Perth in Western Australia. The interferometer consists of 128 phased array tiles which can be electronically steered. Operating between 50 and 300 MHz the array is used by local universe astronomy, ionosphere studies, solar observations, space situational awareness, SETI, and 21 cm cosmology. Since initial construction in 2013 the array has been upgraded in three phases. See \citet{Tingay} for a more detailed description of the array.

\begin{figure}
\centering
\includegraphics[width=\columnwidth]{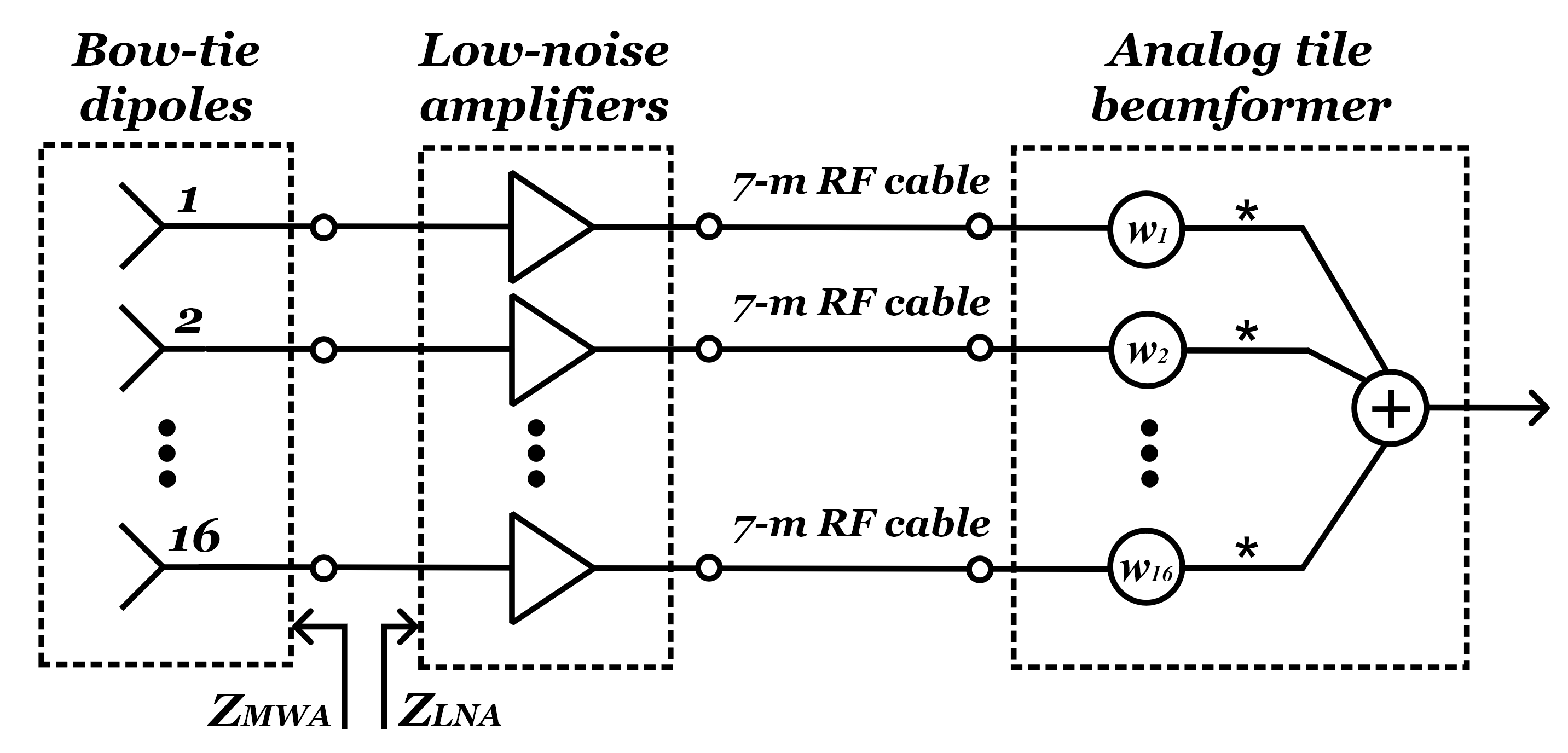}
\caption{Simplified MWA tile system. The output voltage is a scalar. \citep[Image credit:][]{Warnick2018}}
\label{fig:schematic}
\end{figure}
 
An MWA tile, shown schematically in Fig. \ref{fig:schematic} is made up of 16 dual-polarized  antennas arranged in a $4\times4$ regular grid with 1.1~m spacing between antenna centers and placed on a $5\times5$~m conductive ground mesh. Each antenna consists of two orthogonally-mounted aluminum broadband bow-tie dipoles holding custom-designed low-noise amplifier (LNA) circuits at the center. 
The LNA circuit powered from a 5-VDC bias supplied by the analog beamformer includes baluns to convert balanced input to single-ended output. 
Dipoles and tiles are aligned with the East-West (EW) and North-South (NS) directions. The complex voltages from 16 dipoles (32 channels due to dual polarization) are passed through selectable delay lines and added together \citep[][]{Beamformer}. To steer the tile beam, the beamformer applies phase delays to the individual antenna channels. Each beamformer handles both EW (X-) and NS (Y-) linear polarizations. 
 
\begin{figure}
\centering
\includegraphics[width=\columnwidth]{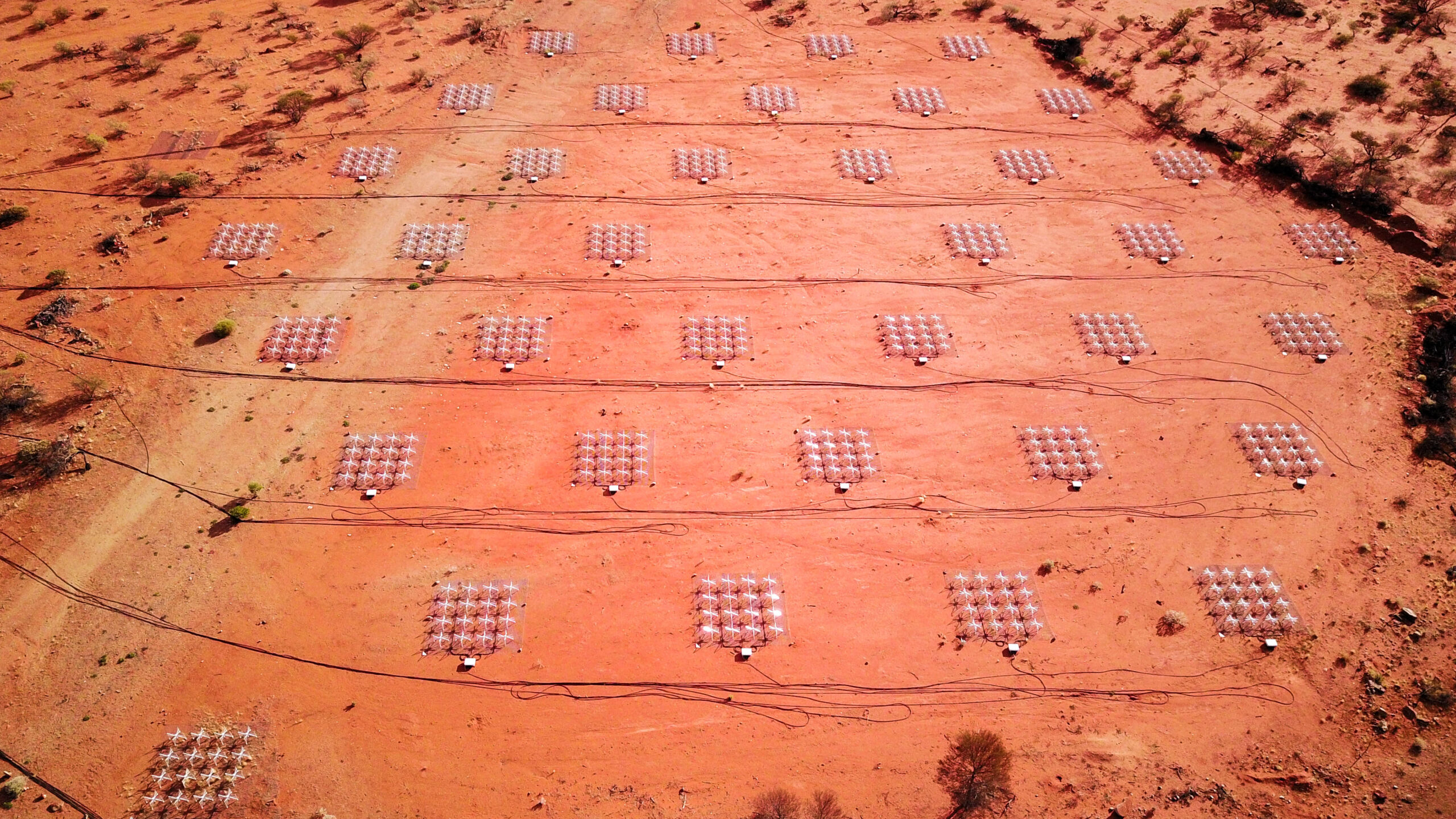}
\caption{MWA Phase II hexagonal subarray as seen from above. Each tile is made up of a $4\times4$ grid of individual bowtie dipoles on a $5\times5$~m ground plane (Image credit: MWA Project, Curtin University).}
\label{fig:phase2}
\end{figure}

The Phase II upgrade of the MWA \citep[][]{MWAPhase2} expanded the array by adding 128 tiles, including 72 tiles arranged in two regular hexagonal configurations near the existing MWA core. Figure~\ref{fig:phase2} is an overhead photo of one of the hexagonal subarrays. The 14~m spacing and hexagonal layout mirror the positions of the HERA telescope dishes. Redundancy, first tested for 21 cm purposes by the PAPER experiment, provides higher power spectrum sensitivity in power spectrum measurements in the limit of lower antenna count \citep[][]{Parsons2012}. \citet{MWAPhase2} reports improved sensitivity on large scales and reduced foreground contamination with the Phase II array power spectrum.

\citet{Kolopanis2023} analyzed data from Phase II of the MWA using the \verb|simpleDS| delay spectrum pipeline. This analysis focused on the three shortest baselines available in the hexagonal subarray, i.e. 14~m, 24.24~m, and 28~m. Short baselines have been excluded from previous analysis because their inclusion was found to increase calibration error. However, short baselines are the most sensitive to the Fourier modes expected of the EoR 21 cm signal and are the most commonly occurring baselines within the hexagonal subarray. 

Inspecting individual baselines, a common mode-type systematic was observed. Excess correlation at delays higher than expected from sky structure was observed to be largely stable across 30 minutes of time but varied widely between nominally redundant baselines. Having established a quality threshold for allowable excess, nearly 50\% of all the data from the shortest baselines were found to exceed it. This included 100\% of the EW baselines in the data set. Even with these most extreme samples excluded, a power spectrum analysis of the 28~m baselines at redshift 7.14 found a statistically significant detection well above previously established limits. Furthermore, the cross power between different baselines was found to be negative, further suggesting a systematic origin. 

\citet{Kolopanis2023} offered theories on the possible source of this systematic. One possible mechanism is unintended radiation of a broadband signal from somewhere nearby. This would show a dependence on location within the array in both amplitude and phase. Another possible mechanism is the sky signal re-radiated to the surrounding antennas from a broken cable as in \citet{Kern2019}. These are both malfunctions in the array operation which might be expected to be fixed with time. A third possibility worth considering because it affects the array as designed is mutual coupling between tiles. 

\section{Mutual Coupling}\label{sec:coupling}

\begin{figure}
\centering
\includegraphics[width=0.8\columnwidth]{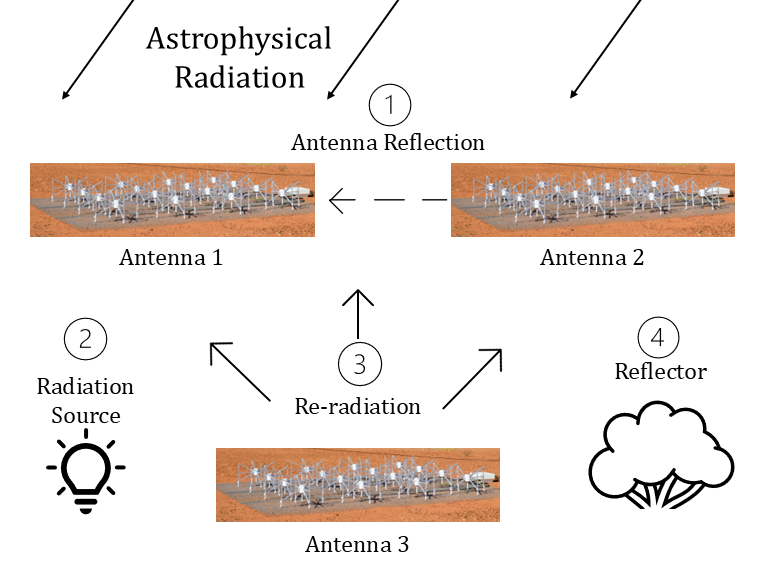}
\caption{Diagram showing various mechanisms of excess correlation in the visibilities. 1) Signal reflecting off of nearby antennas. 2) A broken cable which allows received sky signal to re-radiate. This causes copies of visibilities to appear in other baselines. 3) Impedance mismatch at the antenna feed causes an antenna to re-radiate the sky signal to the surrounding antennas. 4) An object which is reflecting radiation from the sky back into the array. This can be nearby vegetation, equipment, buildings, etc.}
\label{fig:coupling_cartoon}
\end{figure}

In the study of dense phased array antenna systems, mutual coupling often refers to the distortion of individual embedded element patterns (EEPs) resulting from antennas interacting when in close proximity \citep[][]{Kyriakou2025}. While such models are the gold standard, pattern measurements are not always available and large simulations remain a challenging compute problem. A simplified physical situations which could account for a useful fraction of the total coupling could be applied to existing data.

In an interferometer measuring correlation between antennas, coupling between antennas along alternate paths adds to the ``ideal" visibilities. Figure~\ref{fig:coupling_cartoon} illustrates an incomplete list of the possible sources of excess correlation. These include reflections from nearby antennas or objects; re-radiation from broken connections; and re-radiation from neighboring antennas. In the re-radiation model, astrophysical radiation hits all antennas in the array and much of that radiation is absorbed by the antenna. Due to impedance mismatch at the terminals of the antenna feed, some of the radiation is reflected and re-radiated. This radiation is then absorbed by the surrounding antennas --``single path re-radiation''-- or it is reflected off other antennas before being absorbed in a  ``multi-path re-radiation''. 

\citet{Kern2019} modeled mutual coupling in HERA as one antenna's voltage with a coupling coefficient being added to another antenna's voltage before correlation. This two antenna model was sufficiently descriptive of reality to help develop a filtering technique which reduced amplitude by roughly two orders of magnitude in the power spectrum. While this was an improvement, it was still higher than power spectrum estimates require. \citet{Josaitis2022} derived a more complex model which accounts for a third antenna re-radiating and coupling into a visibility. \citet{Rath2024} further developed this model comparing in detail to data from HERA.  The model's prediction of baselines mixing in with delay offsets was found to be in good qualitative agreement in detail in delay and fringe-rate space. However the model was found to under-estimate coupling amplitude by an order of magnitude. Hypothetical causes of this mis-estimation include an under-estimate of the simulated HERA dish horizon gain, multi-path, or a reflection not included in the re-radiation model. 

In this paper we begin to address the question of mutual coupling in the MWA with a first estimate the order of magnitude of the coupling and, by building up from two to three to 21 tiles, examine the effect of embedding and multi-path.

\section{MWA Numerical Model} \label{sec:model}

\begin{figure}
\centering
\includegraphics[width=0.9\columnwidth]{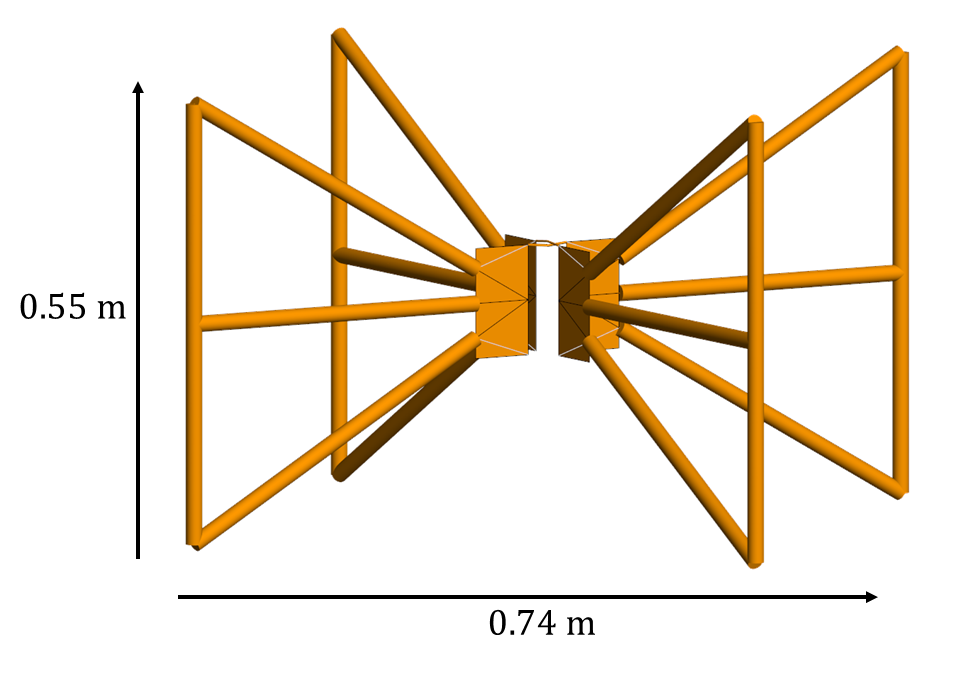}
\caption{The bow-tie MWA dipole antenna modeled in FEKO after \citet{Sutinjo2015Soil}. In keeping with previous models, wire has been substituted for the C-channel used in the as-built antenna.}
\label{fig:feko_dipole}
\end{figure}

\begin{figure}
\centering
\includegraphics[width=\columnwidth]{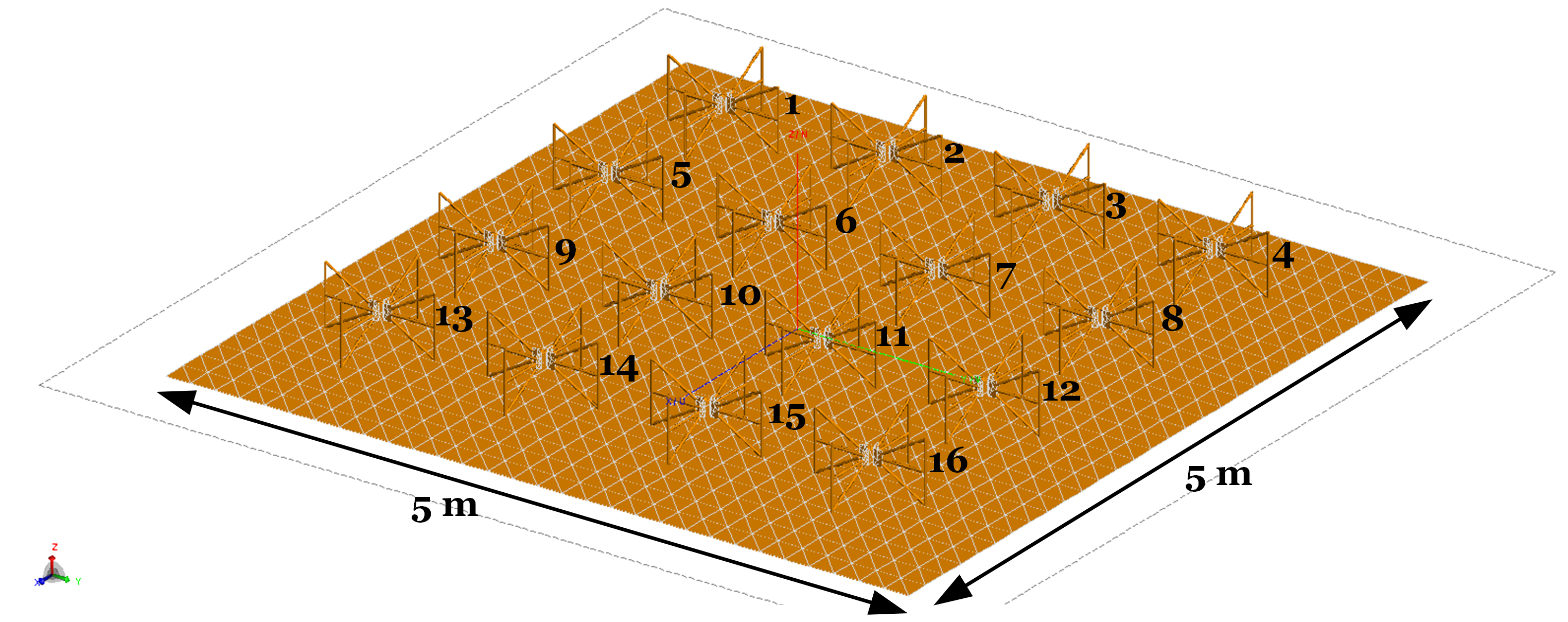}
\caption{A single tile modeled in FEKO. The tile is made up of 16 bow-tie dipoles with 1.1 m spacing. The ground plane is a $5\times5$ m perfect electric conductor (PEC) and the surrounding soil is modeled using permittivity and conductivity measured in-situ \citep[][]{Sutinjo2015Soil}}
\label{fig:tile_model}
\end{figure}

Digital-twin models of the MWA tiles were built in Altair FEKO\footnote{\url{www.feko.info}}, a numerical electromagnetic (EM) simulation software package. The model is a reproduction of the original model used by the MWA collaboration \citep[][]{Sokolowski2017} and shown in Figures~\ref{fig:feko_dipole} and \ref{fig:tile_model}. The FEKO geometry model for the bow-tie dipole has the same dimensions as the physical dipole, but uses 15~mm diameter wires rather than aluminum channels to simplify the simulation. Antennas and the $5\times5$~m ground plane are modeled as perfect electric conductors (PECs). Frequency-dependent soil properties are included using the Sommerfeld formulation in FEKO, assuming a homogeneous, infinite dielectric ground plane. The simulation uses previously characterized soil parameters (relative permittivity and conductivity) from the Murchison Radio-astronomy Observatory \citep[][]{Sutinjo2015Soil}. The interface between the antenna and the LNA is particularly important in the context of the re-radiation model. The LNA has been modeled as a load parallel to the dipole arms with frequency-dependent impedance that was input to the FEKO model as a Touchstone file containing a list of measured impedance values \citep[][]{DISO}.

The EM solver divides this mechanical and electrical model using a tesselating mesh and solves Maxwells equations on this grid under approximations of uniformity within grid cells and continuity across boundaries. Several methods have been devised to solve these equations written in their integral or derivative forms, or by transforming from time to frequency. Each offer tradeoffs in compute time and accuracy. In a study of several simulators and solvers as applied to a 21 cm global experiment, the Method of Moments (MoM) in FEKO was found to provide the most robust model, showing the fewest spurious changes under small parameter variation \citep[][]{Mahesh2021}. \citet{Mahesh2021} found that MoM offers higher accuracy and numerical stability for electrically large PEC structures, minimizing non-physical artifacts due to discretization or meshing errors. Its integral-equation-based approach also inherently accounts for radiation boundary conditions, making it especially robust for antenna modeling. The MoM solver is used for all simulations reported here. Meshing is done using FEKO's automated routine with mesh count increased until the model converged.

\begin{figure}
\centering
\includegraphics[width=\columnwidth]{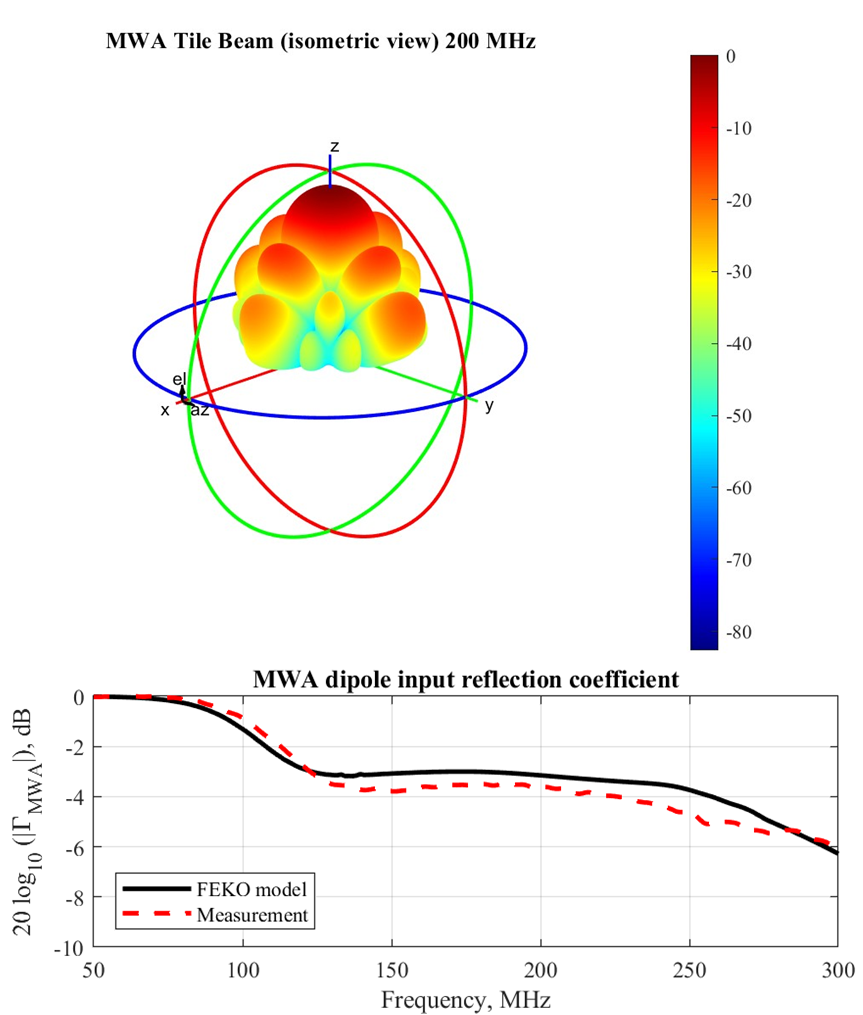}
\caption{(Top) Simulated MWA beam pattern at 200 MHz and (Bottom)~simulated and measured input reflection coefficient of MWA dipole ($Z_0=100\Omega$). It can be seen that the numerical FEKO model matches well the measured response of the antenna.}
\label{fig:gamma_mwa}
\end{figure}

First, simulations with two and three tiles were used to generate scattering parameters. A two tile simulation will provide a baseline cross-coupling between the tiles to be used a reference for comparing higher order effects.  Three tiles are the bare minimum needed to see reflection terms.
The single-tile model with 16 bow-tie dipoles on a finite ground is shown in Figure~\ref{fig:tile_model}. Its scattering parameters and far-field beam pattern are shown in Figure~\ref{fig:gamma_mwa}. These simulations were all run on a high end desktop with 22 cores and 128~GB of ram. A full listing of  computing requirements and run times is given in Table \ref{table:pc_req}. 
The two- and three-tile models finished in 9 and 19 hours respectively. Models with more than three tiles failed to run on the desktop machine with memory demand exceeding that available. 
The 21-tile simulation with a total of 336 dipoles in a half-hexagonal configuration was computed using a 28 cores computer with 1.50~TB of RAM. It required 176 hours to compute S-parameters at 5 frequency points. Simulations with more frequency points will require more powerful high performance computing (HPC) services, as will models with more tiles. Therefore, modeling a full 36-tile hexagonal array is left for future work. Table~\ref{table:pc_req} summarizes the computational requirements for accurate electromagnetic analysis of our models, highlighting the high computational cost involved. The mesh data, given in terms of the number of segments (S) and triangles (T) used to discretize the geometry, demonstrates that large models incur significantly higher peak memory usage not available on a typical machine and require HPC.

\begin{table}
\caption{Computing requirements for electromagnetic simulation of mutual coupling using S-parameters in FEKO.}
\label{table:pc_req}
\resizebox{\columnwidth}{!}{%
\begin{tabular}{cccccc}
\hline
Model   & \begin{tabular}[c]{@{}c@{}}Frequency\\ points\end{tabular} & \begin{tabular}[c]{@{}c@{}}Simulation\\ output\end{tabular}    & \begin{tabular}[c]{@{}c@{}}Machine and\\ peak RAM\end{tabular} & Mesh info                                                      & \begin{tabular}[c]{@{}c@{}}Time, \\ hours\end{tabular} \\ \hline
2-tile  & \begin{tabular}[c]{@{}c@{}}221\\ (80-300 MHz)\end{tabular} & \begin{tabular}[c]{@{}c@{}}.s32p\\ (Y-pol)\end{tabular}        & \begin{tabular}[c]{@{}c@{}}*\\ 281 MB\end{tabular}             & \begin{tabular}[c]{@{}c@{}}S:  1,792\\ T:  2,572\end{tabular}                                                              & 9                                                      \\ \hline
3-tile  & \begin{tabular}[c]{@{}c@{}}221\\ (80-300 MHz)\end{tabular} & \begin{tabular}[c]{@{}c@{}}.s48p\\ (Y-pol)\end{tabular}         & \begin{tabular}[c]{@{}c@{}}*\\ 568 MB\end{tabular}             & \begin{tabular}[c]{@{}c@{}}S:  2,688\\ T:  3,858\end{tabular}                             & 19                                                   \\ \hline
21-tile & \begin{tabular}[c]{@{}c@{}}5\\ (100-300 MHz)\end{tabular}  & \begin{tabular}[c]{@{}c@{}}.s672p\\ (X-pol, Y-pol)\end{tabular} & \begin{tabular}[c]{@{}c@{}}**\\ 1.148 TB\end{tabular}          & \begin{tabular}[c]{@{}c@{}}S: 31,584\\ T: 170,016\end{tabular} & 176                                                    \\ \hline
\end{tabular}%
}
* Intel Core i7-6800K CPU @3.40GHz, 11 parallel processes on 2 CPUs\\
** Intel Xeon Platinum 8180 @2.50GHz; 55 parallel processes on 2 CPUs
\end{table}

\begin{figure}
\centering
\includegraphics[width=\columnwidth]{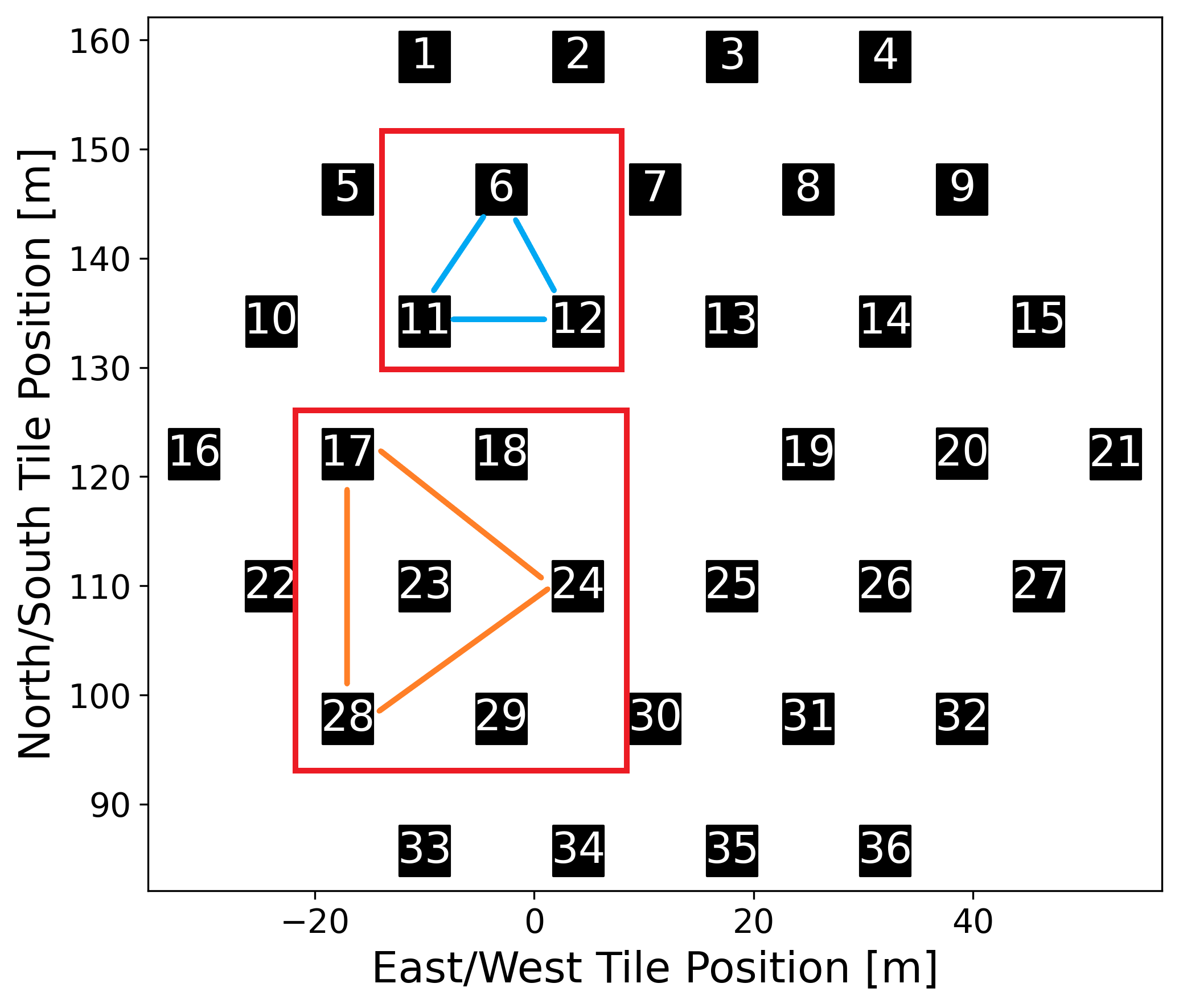}
\caption{The two redundant baseline vectors used in this analysis plotted over a map of one of the hexagonal arrays. The blue baselines at the top are 14~m, and the orange baselines below are 24.24~m.}
\label{fig:tile_map}
\end{figure}

Motivated to explain observations of excess correlation on the shortest baselines reported by \citet{Kolopanis2023} we built simulations of the 14~m and 24.24~m baselines. These are the two shortest and therefore most redundant baselines in the hexagonal subarray and also likely to be the largest coupling. These baselines were built in FEKO, first as two tiles of each baseline length and then as three tiles making up a triangle with the baseline length. Figure~\ref{fig:tile_map} shows a map of the hexagonal array with the two modeled baselines highlighted. 

\section{Simulating Re-Radiation} \label{sec:calc_vari}
The analytic re-radiation model describes a coupling between two antennas according to a geometric link budget \citep[][]{Josaitis2022}. To our knowledge, though re-radiation has been compared against data in HERA \citep{Rath2024}, it has not yet been compared against EM models, and has not been applied to other telescopes. If it is to be a baseline for developing a parametrized physical model of mutual coupling in the MWA in the future, a basis for comparison with EM simulations must be devised.

The relevant comparison point against an EM simulation is the S-parameter. S-parameters describe the complex gain factor between two ports in an electrical system. However, the $S_{12}$ simulation will also include factors not in the analytic model such as reflection from nearby objects, deviations from far field, resonances between antennas, and non-far-field fringing. This will help to inform if there are additional factors which must be accounted for in a parametrized physical model beyond what was derived in \citet{Josaitis2022}.

\subsection{Single Baseline Mutual Coupling and Re-Broadcast Check}\label{sec:2tile_sparam}

A single pair of tiles with a 14~m EW baseline provides no opportunity for multi-path and, at 150 MHz, a 4.4~m tile is reasonably beyond the far field limit at 14~m.\footnote{The far field is at $D=1.22*\frac{D^2_{\mathrm{tile}}}{\lambda} = 11.8$ m}.

\begin{figure}
\centering
\includegraphics[width=\columnwidth]{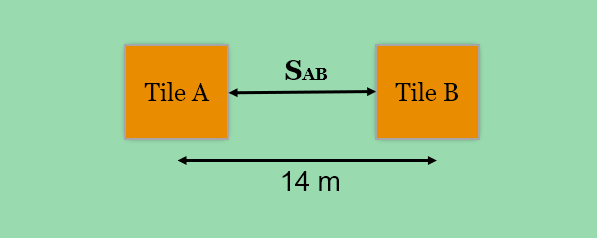}
\caption{The layout of two MWA tiles along the shortest baseline in the hex sub-array separated by 14~m along the EW direction. }
\label{fig:2tile14m}
\end{figure}

Figure~\ref{fig:2tile14m} shows a model of two tiles with a baseline along the East-West direction separated by 14~m, for which the complex scattering parameters between every dipole were simulated using FEKO Method of Moments (MoM). FEKO output $S_{ij}$ between every dipole in the two-tile model for the NS polarization, where $i$ indexes dipole number in Tile A, and $j$ in Tile B. This gives a 32-by-32 S-parameter matrix (1024 complex $S_{ij}$ values for each 1.28~MHz coarse channel) that was calculated for 221 frequency points in the range of 80--300~MHz. The total number of data points is thus 226,304, which is comparable to S-parameter data of dual-polarized SKA-Low station simulated at 3 frequency points \citep[][]{bolliDD2022}. Simulation time of this MWA model is much shorter than that of SKA-Low due to the complexity of antenna structure.

\begin{figure}
\centering
\includegraphics[width=\columnwidth]{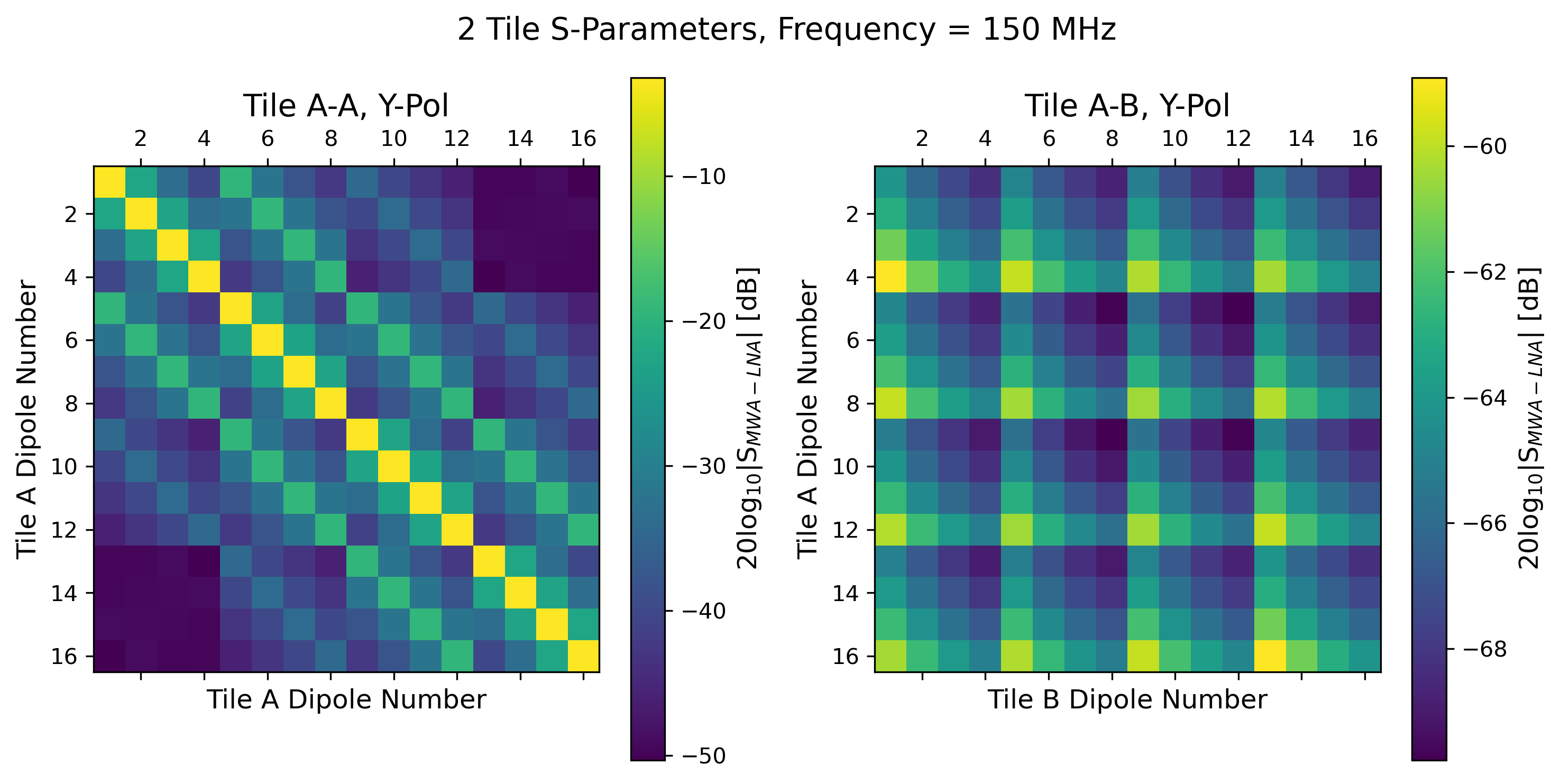}
\caption{Magnitude of the S-parameters between dipoles in two tiles, A and B, with a 14~m EW baseline at 150 MHz. Due to reciprocity, S-parameters of Tile A are equivalent to Tile B, and therefore, Tile B S-parameters are not shown. In the Tile A plots, the diagonal corresponds to $S_{ii}$ and block diagonal portions corresponds to coupling within each tile. The magnitude is lower for larger separation between dipoles. The x- and y-axis labels both correspond to the dipole numbers from Fig \ref{fig:tile_model}. The grid pattern of coupling between tiles reflects the distance with closer antennas having higher coupling.}
\label{fig:sparam_matrix}
\end{figure}

To accurately model mutual coupling, the  reference plane must be between the antenna and the LNA, and then sum all signals within a tile to simulate a "single" antenna.
To de-embed the LNA, we converted the S-parameters from the FEKO simulation to impedance, subtracted the LNA impedance and converted the mutual impedance Z-matrix to a scattering S-matrix \citep[][]{Ung2020}. This brings the reference plane from the LNA output port to the output ports of individual antennas \citep[][]{Ellingson}. Figure~\ref{fig:sparam_matrix} shows the reflection coefficient in Tile A, which is the same as Tile B, as well as the coupling between Tile A and Tile B for NS-polarized dipoles. The diagonals in the Tile A-A matrix correspond to the input reflection coefficient of each dipole, while the off-diagonal elements correspond to the coupling between the dipoles within the tile. The magnitude of the coupling drops off as the distance between dipoles increases, ranging from about -20 to -50~dB. These matrices are symmetric because of the symmetry of the tile.

The matrix representing mutual coupling between Tile~A and Tile~B has a noticeable grid pattern. This is due to the fact the tiles are uniform planar arrays with equal separation between antennas.  Tile~A dipoles closer to Tile~B couple stronger than the dipoles on the far side. The dipoles on the inner edges closest to each other have a magnitude of about -60 to -62 dB, while the dipoles on the outer edges furthest from each other have magnitudes of about -66 to -69 dB. This follows the inverse square law as the distance increases, which confirms the expectation of how mutual coupling behaves.

\begin{figure}
\centering
\includegraphics[width=\columnwidth]{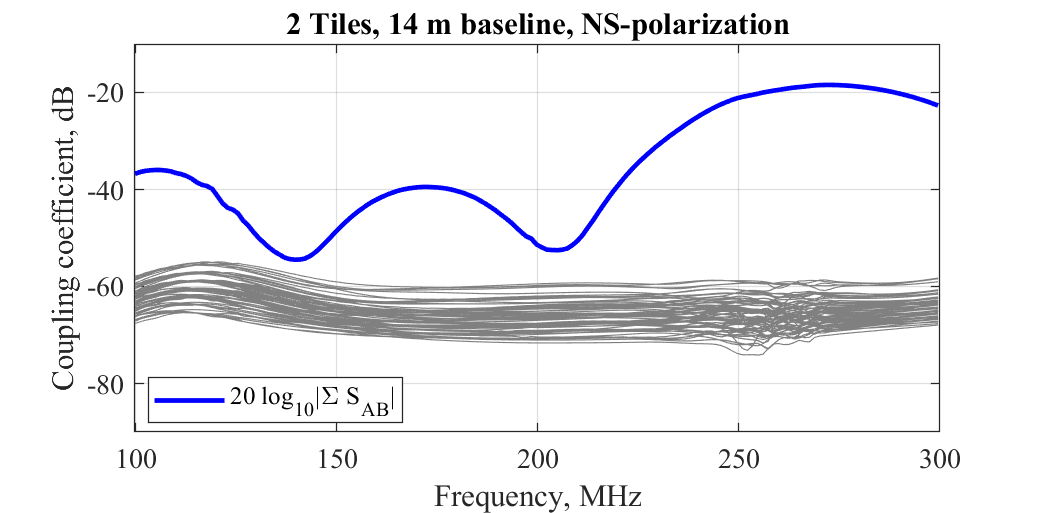}
\caption{Coupling coefficient for a 2-tile model as a function of frequency. The gray lines are the magnitude of S-parameters for every dipole pair between tiles A and B. The green curve is the sum of all $S_{ij}$'s.}
\label{fig:sparams}
\end{figure}

The MWA beamformer sums the analog signals from all the 16 dipoles for each polarization. The net coupling between each tile is the sum the S-parameters across antennas within each tile.
Figure~\ref{fig:sparams} plots $S_{AB}$, the summed S-parameters between tiles A and B. $S_{AB}$ and $S_{BA}$ are equal because of the symmetry of the tiles and therefore $S_{BA}$ has not been plotted. If added in phase, the sum over all pairs of dipoles would increase by $16^2$ or 25~dB. Compared to the individual dipole pairs, the total coupling is modulated as a function of frequency with nulls at 140 and 210 MHz. This spectral structure of the sum follows the beam response in the direction of the baseline.

\subsection{Mutual Coupling in a 3-Tile Model}

\begin{figure}
\centering
\includegraphics[width=0.8\columnwidth]{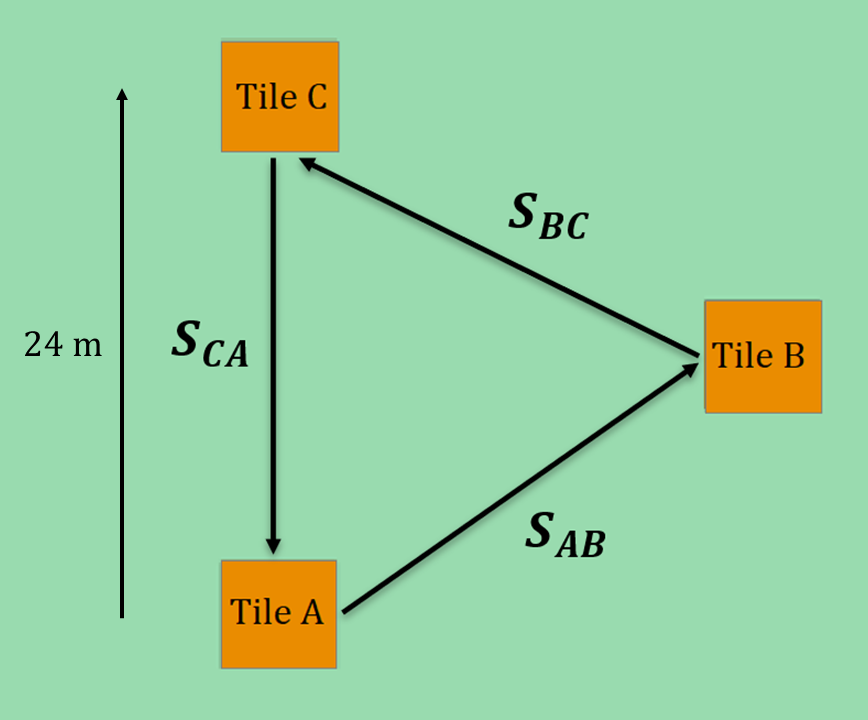}
\caption{The layout for the three tiles with baseline length of 24.24~m. This forms an equilateral triangle with a NS baseline. Each tile is labeled with a letter to differentiate them. The tile internal to the triangle, seen in Figure~\ref{fig:tile_map}, is not simulated here.}
\label{fig:24m}
\end{figure}

\begin{figure}
\centering
\includegraphics[width=\columnwidth]{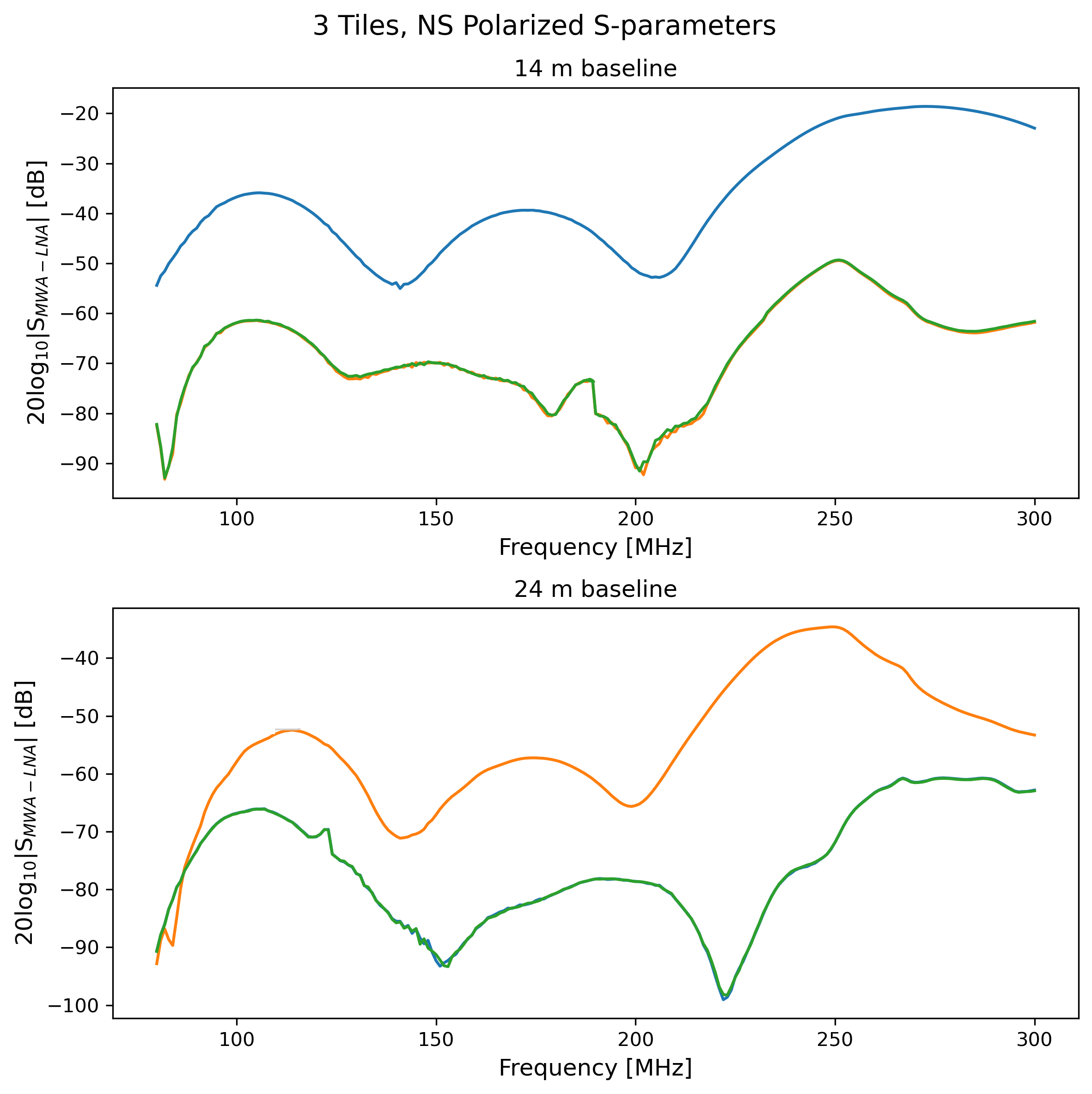}
\caption{$S_{ij}$ simulations of the three tile models, all using NS polarization. Both models produce sums with spectral structure reflecting the beam response in the direction of the baseline. (Top) $S_{AB}$ is the EW baseline and has the highest coupling. The symmetric diagonals $S_{BC}$ and $S_{CA}$ are almost equal. The model tiles were positioned using the real measured locations of the tiles and therefore the distances between tiles are not exactly equal. This may account for the slight difference seen. (Bottom) $S_{CA}$ is the NS baseline and has the highest magnitude of the three. The $S_{AB}$ and $S_{BC}$ measurements are almost equal with slight differences for the same reason as the shorter baseline.}
\label{fig:3tile_sparams}
\end{figure}

This same analysis was repeated for the NS polarized three tile models and produced similar  matrices and $S_{ii}$ parameters. The matrix plots are available in Appendix~\ref{sec:appendix_3tiles}. Figures \ref{fig:14m} and \ref{fig:24m} show the layout for the two models, with labels indicating the tile names and corresponding S-parameter in Figure~\ref{fig:3tile_sparams}. Both models produce sums with spectral structure reflecting the beam response in the direction of the baseline. The 14~m baselines have a higher amplitude, which is to be expected as the tiles are closer together and the EW baseline means the NS polarized antennas are aligned. In both models, the diagonal baselines have lower amplitudes than the horizontal and vertical baselines. 

\begin{figure}
\centering
\includegraphics[width=0.8\columnwidth]{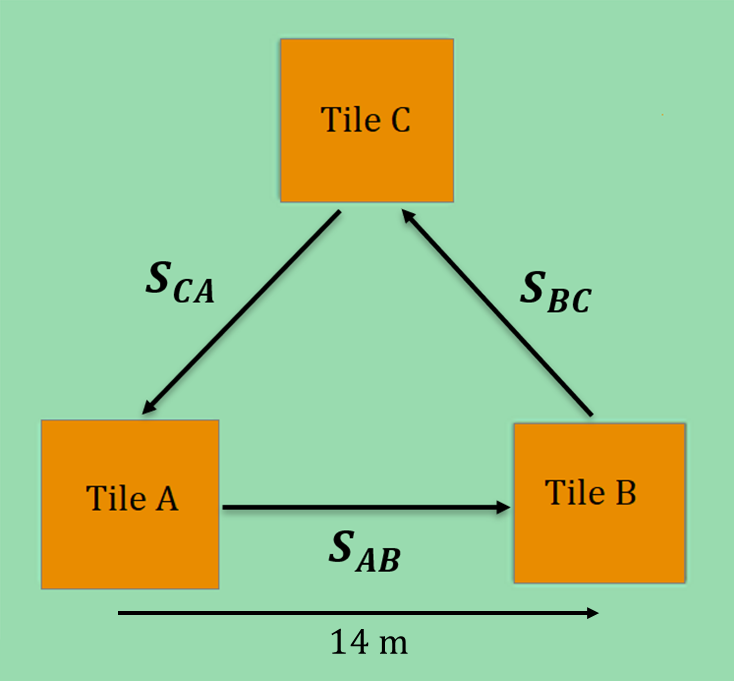}
\caption{The layout for the three tiles with baseline length of 14~m. This forms an equilateral triangle with an EW baseline. Each tile is labeled with a letter to differentiate them.}
\label{fig:14m}
\end{figure}

\subsection{Mutual Coupling in a 21-Tile Model}

\begin{figure}
\centering
\includegraphics[width=\columnwidth]{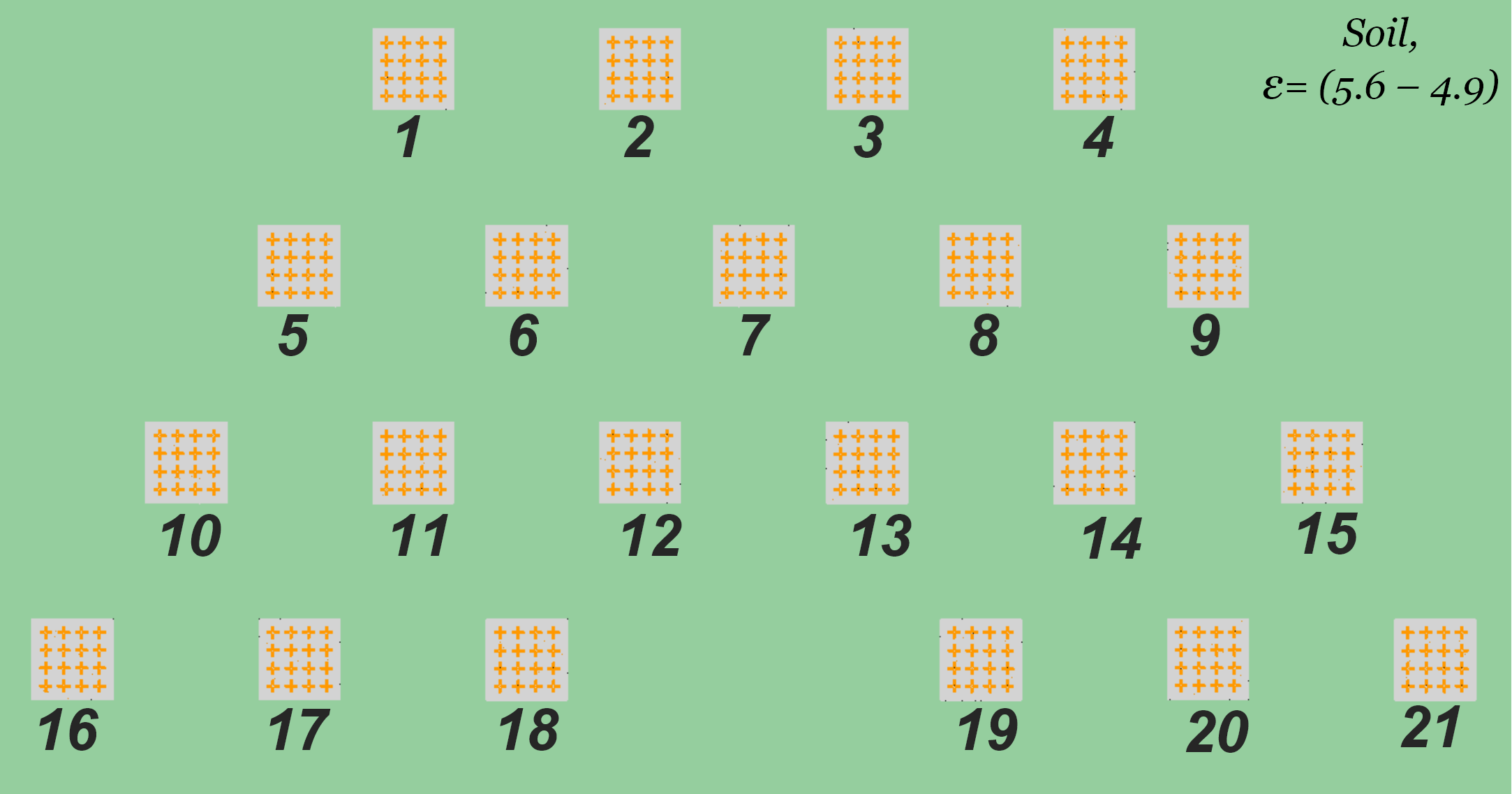}
\caption{The layout for the 21-tile model. The two rows most distant from the reference antennas have been removed to make the model fit into the best-available 1.5TB of RAM.}
\label{fig:21tile_model}
\end{figure}

\begin{figure}
\centering
\includegraphics[width=\columnwidth]{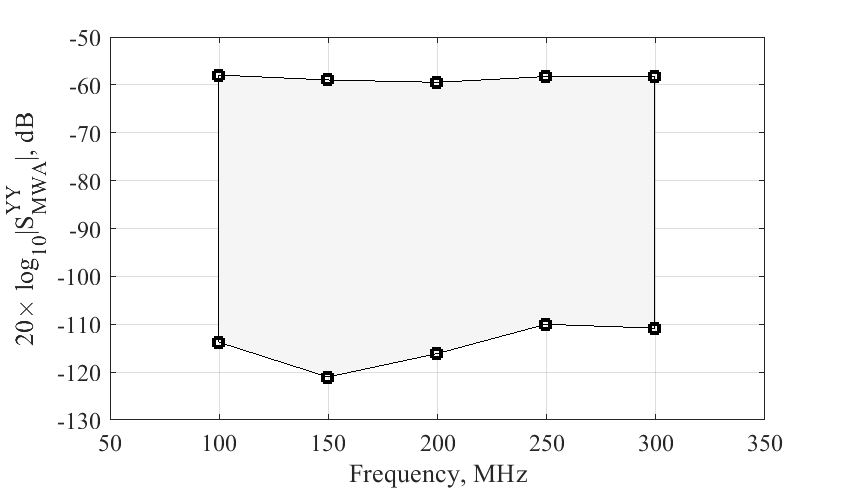}
\caption{The minimum and maximum range of dipole-dipole scattering parameters in a simulation of 21 tiles (Figure~\ref{fig:21tile_model}) across 112896 pairs of Y-polarized antennas (NS).}
\label{fig:21tile_sparams}
\end{figure}

With the intention to obtain mutual coupling between MWA tiles in a hexagonal configuration though the electromagnetic simulation, we created a model of a hexagonal array in FEKO. The most powerful machine available to us was an Intel Xeon Platinum 8180 CPU@2.50GHz with 28 cores and 1.50~TB RAM. The memory demand for the 36-tile hexagonal array exceeded the available RAM, so the array was reduced to 21-tiles by removing the rows of antennas most distant from the reference baselines and reducing the number of frequency channels. A map of this model is shown in Figure~\ref{fig:21tile_model}. This model consists of 31,584 mesh PEC segments and 170,016 mesh PEC triangles in FEKO. This simulation took 176~hours to complete for 5 frequency points on the aforementioned machine. Figure~\ref{fig:21tile_sparams} plots the envelope of the maximum and minimum dipole-dipole S-parameters for the 21-tile model. The maximum values fall in the same range as those shown in Figure~\ref{fig:sparams} for the 2-tile model, however the minimum values are much lower due to the much longer baselines available between dipoles. 

\begin{figure}
\centering
\includegraphics[width=\columnwidth]{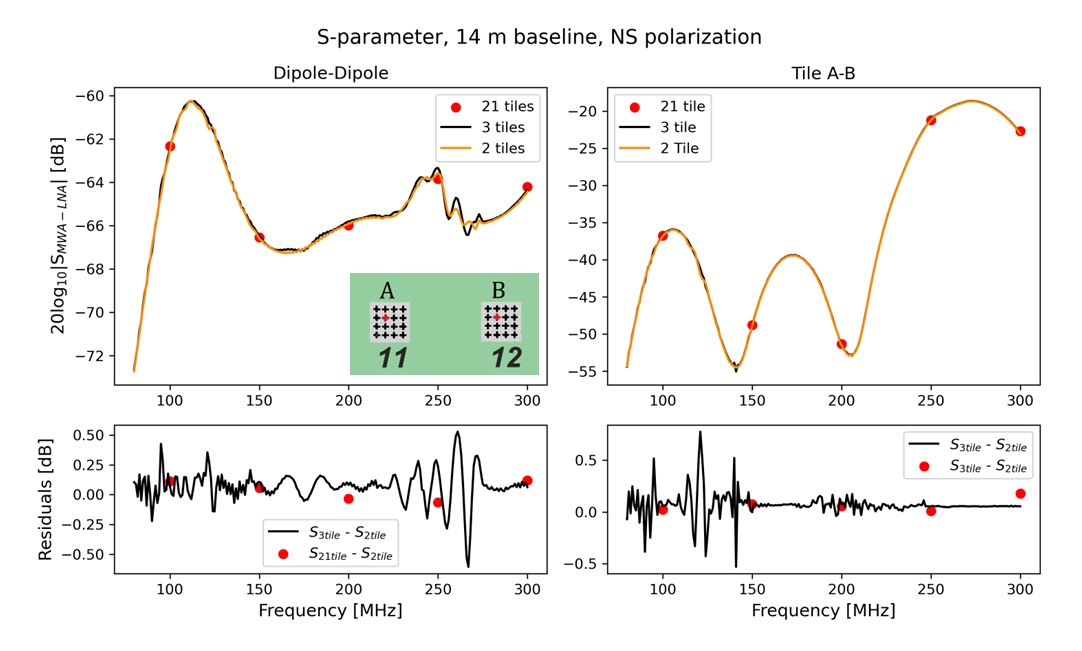}
\caption{A comparison of the S-parameters for the different 14~m baseline FEKO models. The 21-tile model, plotted as red dots, only has 5 frequency points. (Top Left) All three models are plotted for a single dipole-dipole baseline, which is highlighted in the map inset. The three models are very similar. (Bottom Left) The residuals after subtracting the two tile model from the three tile and 21-tile models. This highlights the difference in spectral structure between the models. (Top Right) All three models are plotted for a tile-tile baseline. We can again see that the models match almost exactly. (Bottom Right) The residuals after subtracting the two tile model from the three tile and 21-tile models. This again highlights the difference in spectral structure between the models. In particular, there is more difference between the two and three tile models at low frequencies.}
\label{fig:21tile_comparison}
\end{figure}

\begin{figure}
\centering
\includegraphics[width=\columnwidth]{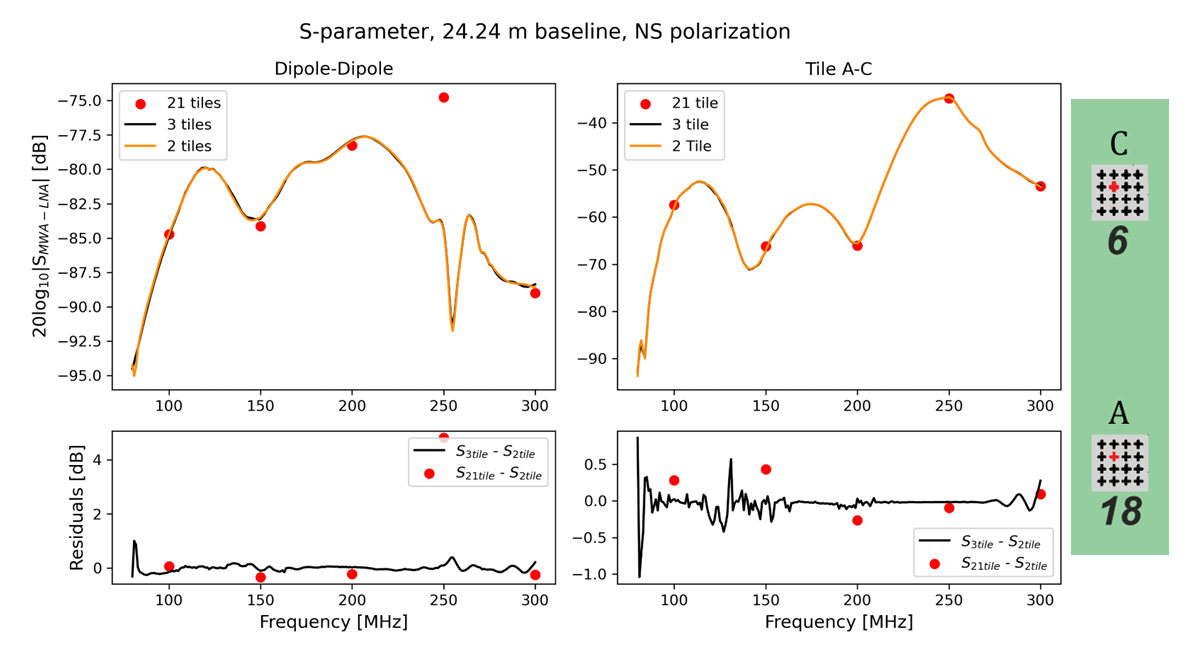}
\caption{A comparison of the S-parameters for the different 24.24~m FEKO models. The 21-tile model, plotted as red dots, only has 5 frequency points. (Top Left) Both models are plotted for a single dipole pair, which is highlighted in the map on the far right. We can see that like the 14~m models in Figure~\ref{fig:21tile_comparison}, the 21-tile model is generally aligned with the other models, except at 250 MHz. (Bottom Left) The residuals after subtracting the two tile model from the three tile and 21-tile models. This highlights the difference in spectral structure between the models. The greatest difference is at 250 MHz. (Top Right) The sum of all $S_{AC}$ for both models are plotted. As with the shorter baseline, the models are very similar. (Bottom right) The residuals after subtracting the two tile model from the three tile and 21-tile models. Compared to the shortest baseline, embedding in a larger 21 tile array results in larger residuals with more spectral structure.}
\label{fig:21tile_24m_comparison}
\end{figure}

Figure~\ref{fig:21tile_comparison} plots a comparison of three FEKO models: two tiles with a 14~m baseline, three tiles with a 14~m baseline, and a 14~m baseline in the 21-tile model. The map set in the plots highlights which tiles and dipoles were selected. The top left plot shows the S-parameter for a single baseline between the two selected dipoles, while the top right plot shows the sum for all $S_{ij}$ between Tiles A and B. Because the 21-tile model only had 5 frequency points, only limited conclusions can be drawn about spectral structure. However, we can see that for both the dipole-dipole and the tile-tile S-parameters, the three models are extremely similar. 
The bottom plots are the residuals left over after subtracting the two tile model from the three tile and 21-tile models. Difference in spectral structure between the sums of the two and three tile models is localized in frequency at the .25 to 0.5 dB level and decreasing towards higher frequencies.

Figure~\ref{fig:21tile_24m_comparison} plots a comparison of three FEKO models: two tiles with a 24.24~m baseline, three tiles with a 24.24~m baseline model, and a 24.24~m baseline in the 21-tile model. The map on the far right of the plot highlights which tiles and dipoles were selected. The top left plot shows the S-parameter for a single baseline between two selected dipoles, while the top right plot shows the sum for all $S_{ij}$ between Tiles A and C. Because the 21-tile model only had 5 frequency points, spectral structure can only be compared on a very coarse scale. The bottom plots are the residuals left over after subtracting the two tile model.

The dipole-dipole S-parameters show general agreement between the models. There is $\sim1$ dB difference, except at 250 MHz, where there is a 5 dB difference in amplitude. The tile-tile S-parameters, shown on the right, show general agreement for all frequencies. The residuals show some difference in spectral structure between the two and three tile models, with the difference $\sim1$ dB.

These results imply that while adding tiles increases the spectral structure, the overall amplitude does not change significantly for models with more than two tiles. The spectral structure added by increasing from three tiles to 21 can't be evaluated until more channels are available. 

\section{Tile Coupling vs Baseline}\label{sec:SvB}

\begin{figure*}
\centering
\includegraphics[width=\textwidth]{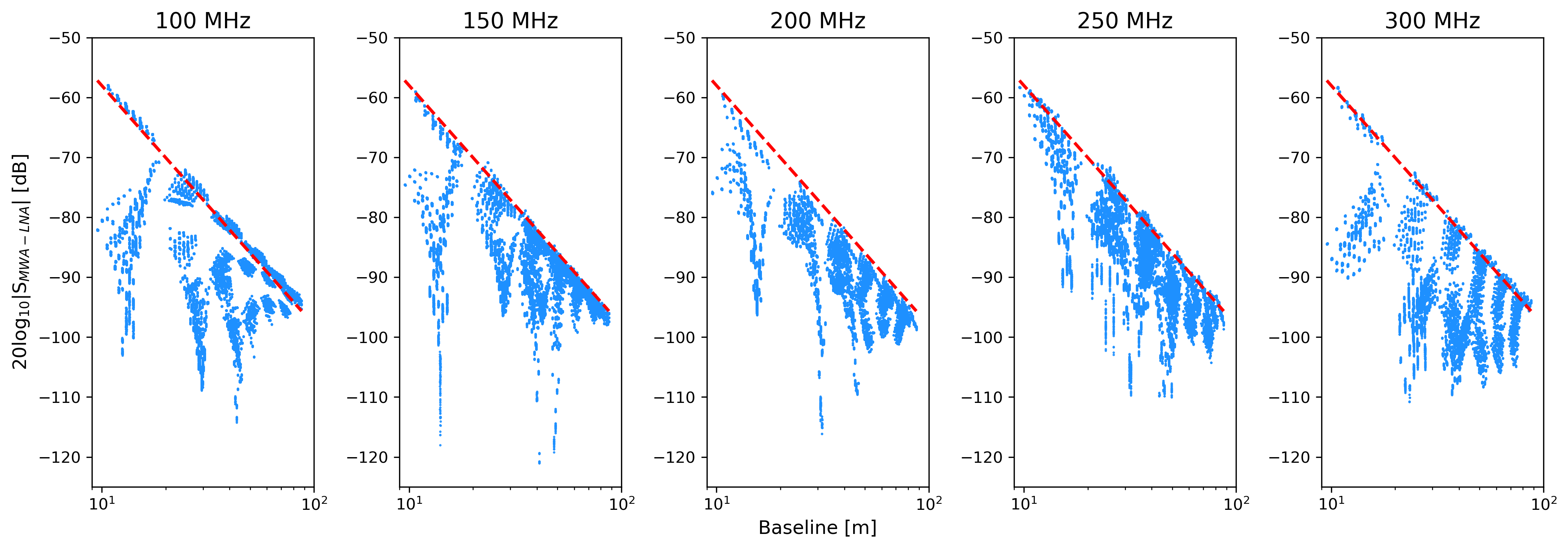}
\caption{The amplitude of the S-parameters of the 21-tile model as a function of baseline length for 5 frequencies. This is plotted in log-log scale. The blue dots are the individual S-parameters. The red dashed line is a scaled $1/r^2$ fit line.}
\label{fig:SvB}
\end{figure*}

In the re-radiation models of mutual coupling, propagation is assumed to follow free-space path loss, with amplitude decreasing inversely with distance $1/b_{ik}$. Figure~\ref{fig:SvB} plots the S-parameters for the 21-tile model as a function of baseline length for 5 frequency points. The beam pattern is visible in the way similar baseline lengths are spread out over a range of amplitudes. 
A scaled second-order power law is plotted over the data. While the beam pattern introduces significant scatter, the general trend of the data follows the power law. This confirms that FEKO follows free-space path loss and is the correct method for an analytic model.

\section{Discussion} \label{sec:discussion}

Mutual coupling between MWA tiles has not been directly studied in the literature or more generally between phased array antenna stations. Numerical models are computationally expensive and time consuming while the level of coupling is expected to be relatively low and perhaps difficult to observe in practice. The
EM simulations presented here predict a level of coupling which present concerns for 21 cm detections. Addition of one antenna into a two antenna system adds spectral structure at the $5$\% level. Addition of many more antennas approaching the true size of the array does seem to introduce even more structure, however it is difficult to estimate the amount due to low frequency resolution. The large array could only be simulated at a small number of channels which limits any observations of the spectral structure of the coupling in the full array.

\subsection{Coupling Requirement for 21 cm Signal Detection} \label{sec:21requirement}

Arguably the most important factor in the impact of coupling is the overall amplitude which sets the subsequent amplitude of the cosmological spectral modes. One way to evaluate whether the amplitude is suitable is to forward model to a power spectrum and compare with forecast 21 cm level. Another complementary approach is to work backwards from forecast 21 cm levels to set a  requirement for the maximum acceptable coupling. 
A simplistic requirement would hold that the worst-case scenario would be for all mutual-coupling power to land in the 21 cm window. Thus the coupling factor must be smaller than the ratio of the foreground power spectrum to the 21 cm signal. Recent limits by \citet{Trott2020} report foreground levels at $1e15$ mK$^2$ and forecasts call for power spectrum levels of 100 mK$^2$. This gives a ratio of $1e14$ in power spectrum, $1e7$ in visibility and $1e3.5$ in voltage. This  corresponds to an allowed coupling of -70 dB\footnote{Units of visibility are also units of power.}. 

\subsection{Limitations and Approximations}

Several approximations have been made here which limit generalization and motivate future work.
The EM simulations used here were kept as small and simple as possible while still showing the effect of ``multi-antenna'' coupling. This allows for only limited multi-path or fringing effects. No surrounding passive environment (such as rocks or plants) were included, nor were any imperfections in the individual dipoles and tiles. Factors like this have been found to perturb the beam and could introduce further non-redundancy and variation \citep[][]{Chokshi2024}. Future investigations may need to include more complex surroundings and imperfections in the array.

Additionally, the estimated level of maximum allowable coupling for 21 cm experiments is at best a crude guide and could be improved. A better estimate of the requirement would take into account the ``wedge'' coupling of foreground power into the 21 cm modes. This would result in a much less strict requirement on coupling, but would apply in specific delay modes where coupling might be relatively stronger. More study is needed in this area. 

\subsection{Parameterized Physical Models}

Though EM simulations are potentially the most accurate word on antenna response, there are good reasons to build a simpler model. Numerical simulations are computationally expensive, as previously discussed. This severely limits exploration of parameter space. Additionally, simulations are not necessarily robust. We have found that with reasonable (but ultimately flawed) configurations of the solver, small changes in design can cause large changes in response. Careful validation by comparison to standard models and between solvers is necessary to build confidence in results \citep[][]{Fagnoni2021,Mahesh2021}. A parametrized model which can be solved and inverted is needed to calibrate effects of mutual coupling in real data. Even if inversion of effects from data is ultimately infeasible, for all of these reasons a simplified physical model will help understand how different effects contribute to the process. 

The re-radiating model derived by \citet{Josaitis2022} is ``semi-analytic'' because it relies on numerical EM simulations of an isolated antenna to determine quantities such as the impedance and the radiation pattern, and then these results are used to calculate the expected coupling between two antennas in the form of a link budget. Mathematically the model works out such that the visibility for baseline $ij$ has added to it copies of baselines between the two participating antennas and every other antenna in the array. \citet{Rath2024} compared this model to observations from HERA and found that the model does indeed reproduce the observed temporal and spectral behavior of HERA data, however it under-predicts the amplitude by roughly an order of magnitude. This means that while the model can be helpful in devising a time-and-frequency filtering scheme to reduce coupling in the EoR window, the model is an incomplete reflection of reality. The reason for model inaccuracy remains unknown and further validation is required. 

Applying this analytic model to the MWA would provide a faster and less computationally expensive way of measuring mutual coupling and filtering observations. However, the MWA is a phased array which uses a beamformer while HERA does not. Further investigation is required to determine how to incorporate the beamformer into the analytic model before the numerical EM model can be compared. 

\section{Conclusion}\label{sec:conclusion}

The original question we sought to answer was how much excess correlation mutual coupling contributes to the MWA and if it was higher than the current EoR power spectrum limits. There are no existing studies on mutual coupling in the Phase II MWA and it was therefore unknown if more thorough research is required.

We built five different models in FEKO increasing in complexity to select for specific coupling effects and proposed a crude coupling requirement motivated by forecast 21 cm levels and known foreground levels. Comparing a model of three antennas to a model with two antennas, it was possible to see spectral structure introduced by a nearby station, presumably through multi-path-like effects, modified the scattering parameters at a $\sim2$\% level. A large simulation of 21 tiles, forming most of one MWA hexagonal subarray found higher levels closer to $5$\% and more uniformly across the band. No conclusions could be drawn regarding the change in spectral structure for more than three tiles due to the frequency resolution of the 21-tile model. 

If somehow mutual coupling were to couple foreground power into a 21 cm line of sight mode, then coupling would need be less than -70 dB in visibility units \citep[][]{Trott2020}. With the exception of the 14~m baselines, this requirement is satisfied below 240 MHz, well above the frequency range for the EoR. The 14~m numerical model exceeds this limit for all frequencies. Mutual coupling could therefore pose a systematic limitation on these shortest baselines. As these are the most abundant baselines in the hexagonal subarray, this is a concern that warrants future research. 
Further investigation into these models is needed before specific mitigation suggestions can be made. 

\section*{Acknowledgments}
This work is supported by the National Science Foundation under grant number AST-2104350.

This scientific work makes use of the Murchison Radio-astronomy Observatory, operated by CSIRO. We acknowledge the Wajarri Yamatji people as the traditional owners of the Observatory site. Support for the operation of the MWA is provided by the Australian Government (NCRIS), under a contract to Curtin University administered by Astronomy Australia Limited. We acknowledge the Pawsey Supercomputing Centre which is supported by the Western Australian and Australian Governments. The MWA Phase II upgrade project was supported by Australian Research Council LIEF grant LE160100031 and the Dunlap Institute for Astronomy and Astrophysics at the University of Toronto.

\section*{Data Availability}

The data products used in this analysis may be requested via email from the authors.



\bibliographystyle{mnras}
\bibliography{main_paper} 




\appendix

\section{3-Tile Models (Two baselines)}\label{sec:appendix_3tiles}

In the interest of completeness, we have included the S-parameter plots for the three tile models. These plots repeat the process described in Section~\ref{sec:2tile_sparam} for the two tile model. 

\begin{figure}
\centering
\includegraphics[width=\columnwidth]{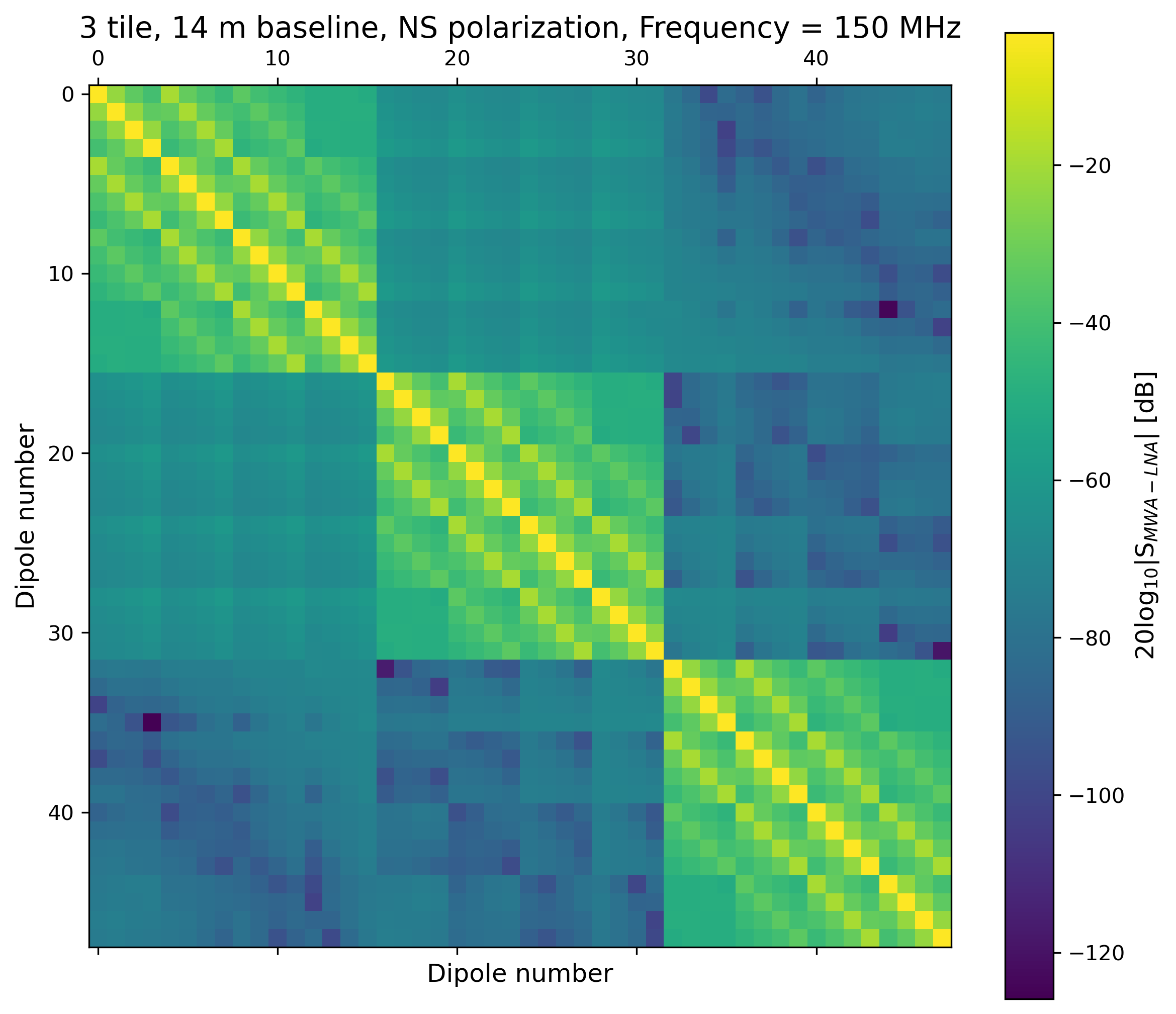}
\caption{The full S-parameter matrix for the three tiles with a baseline of 14~m model.}
\label{fig:3tile_14m_matrix}
\end{figure}

\begin{figure}
\centering
\includegraphics[width=\columnwidth]{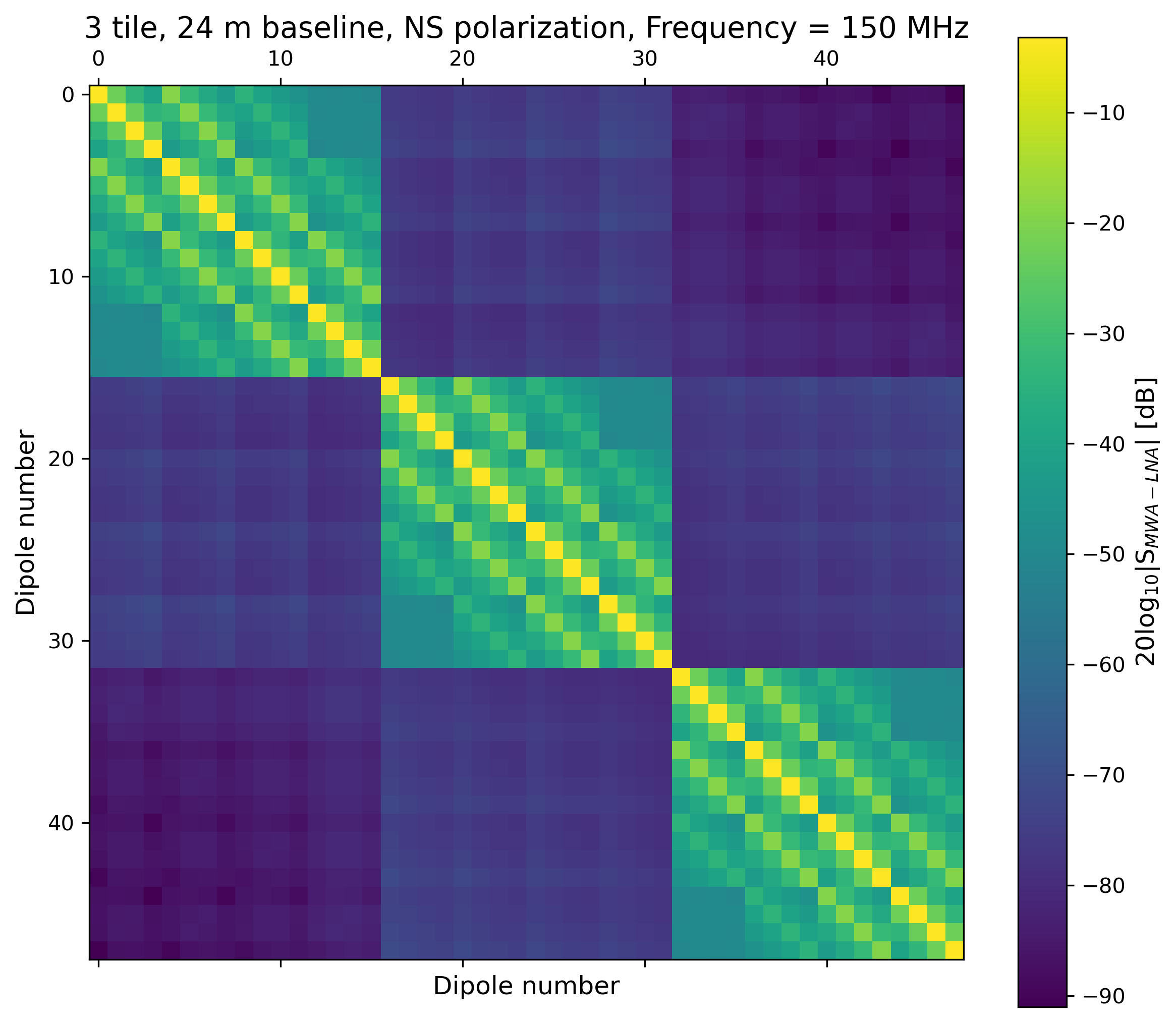}
\caption{The full S-parameter matrix for the three tiles with a baseline of 24.24~m model.}
\label{fig:3tile_24m_matrix}
\end{figure}

Figure~\ref{fig:3tile_14m_matrix} is the full S-parameter matrix for the three tile, 14~m baseline model at 150 MHz. We observe a similar pattern as the two tile matrix in Figure~\ref{fig:sparam_matrix}, with the diagonal representing $S_{ii}$ for each dipole. There are clear $16\times16$ blocks which correspond to the tile-to-tile S-parameters. The three blocks which fall on the diagonal are $S_{AA}$, $S_{BB}$, and $S_{CC}$. The off diagonal blocks are the S-parameters with mutual coupling. 

Figure~\ref{fig:3tile_24m_matrix} is the full S-parameter matrix for the three tile, 24.24~m baseline model at 150 MHz. We observe the same pattern as in Figure~\ref{fig:3tile_14m_matrix}, however the $S_{ij}$ tile pairs have significantly lower amplitudes than the $S_{ii}$ tile pairs. 

\begin{figure}
\centering
\includegraphics[width=\columnwidth]{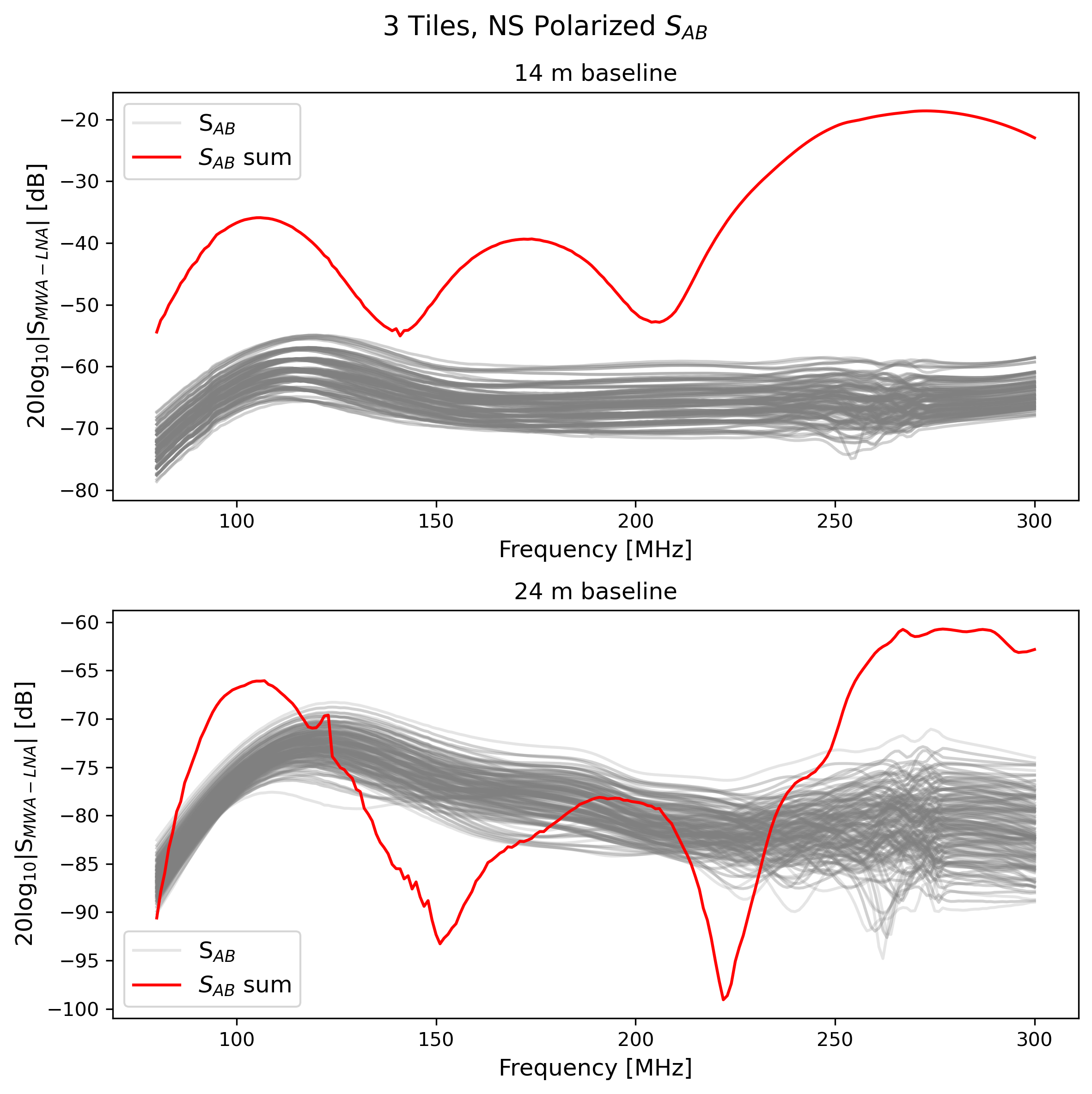}
\caption{Two plots of the magnitude in dB of the S-parameters as a function of frequency. (top) The gray lines are the S-parameters for every dipole pair between Tiles A and B for the 14~m baseline. The red line is the sum of all $S_{AB}$. (bottom) The gray lines are the S-parameters for every dipole pair between Tiles A and B for the 24.24~m baseline. The red curve is the sum of all $S_{AB}$.}
\label{fig:3tile_all_sparam}
\end{figure}

Figure~\ref{fig:3tile_all_sparam} plots all of the S-parameters as a function of frequency for the 14~m baseline in the top plot and the 24.24~m baseline in the bottom plot. Both plots have the sum of all S-parameters plotted in red. 

\section{21-tile model}\label{sec:appendix_21tile}
\begin{figure}
\centering
\includegraphics[width=\columnwidth]{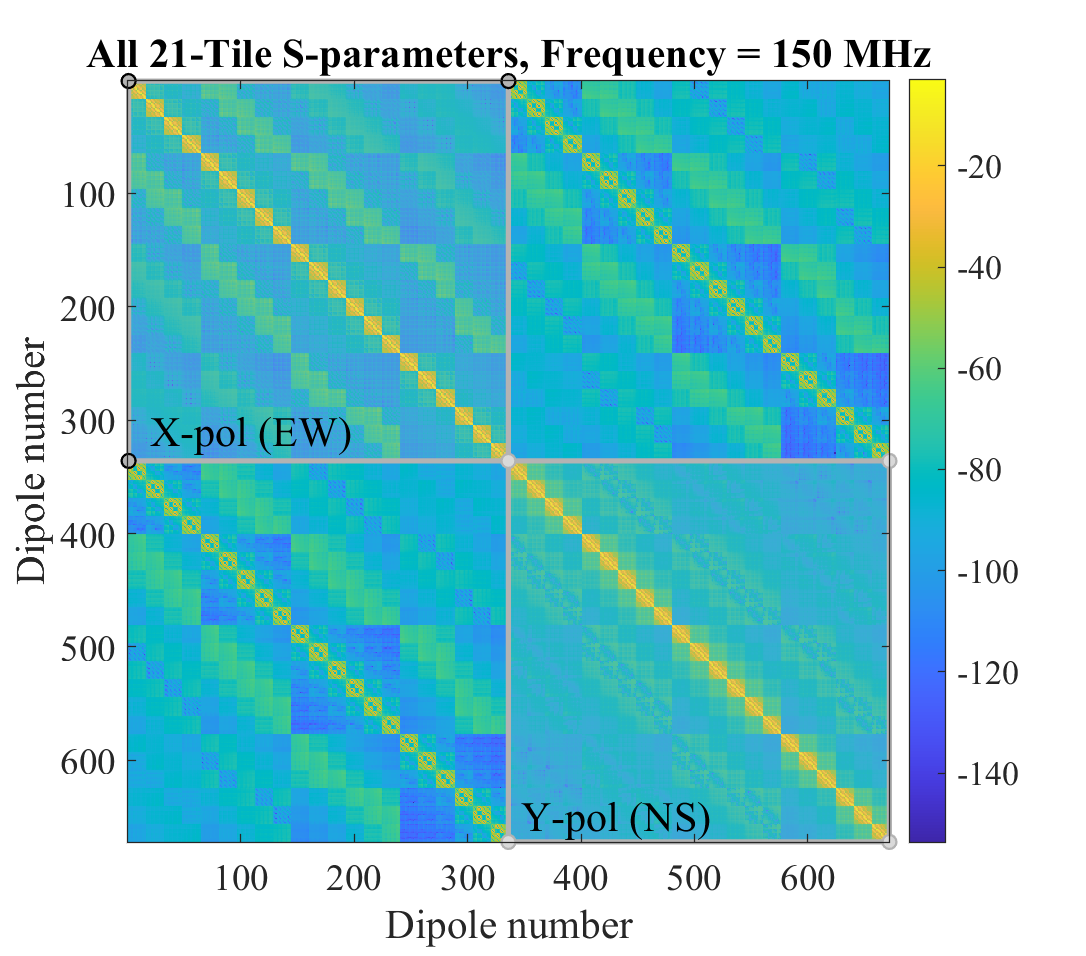}
\caption{Full S-parameter matrix for the 21-tile model, for both polarizations. In this paper, we focused on NS-polarized dipoles (parallel) because they couple 10~dB stronger than EW-polarized dipoles (collinear). }
\label{fig:21tile_matrix_all}
\end{figure}

\begin{figure}
\centering
\includegraphics[width=\columnwidth]{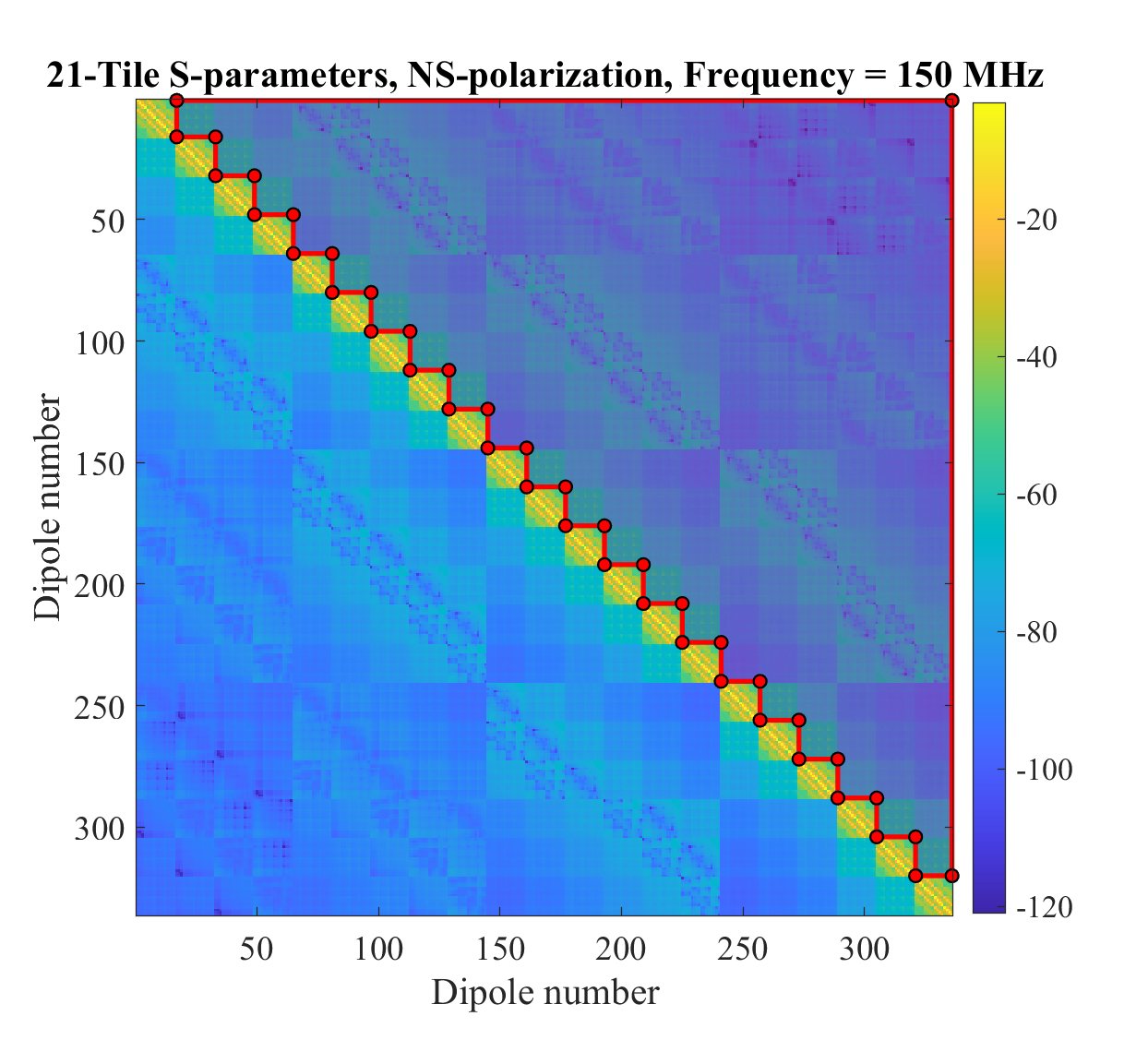}
\caption{S-parameter matrix for the 21-tile model, NS polarization. Diagonal $16\times16$ arrays are the coupling within each tile and antenna self-impedances. For the analysis on inter-tile coupling, we extracted cross-tile mutual coupling coefficients shown within the red polygon. }
\label{fig:21tile_matrix_Ypol}
\end{figure}

\begin{figure}
\centering
\includegraphics[width=\columnwidth]{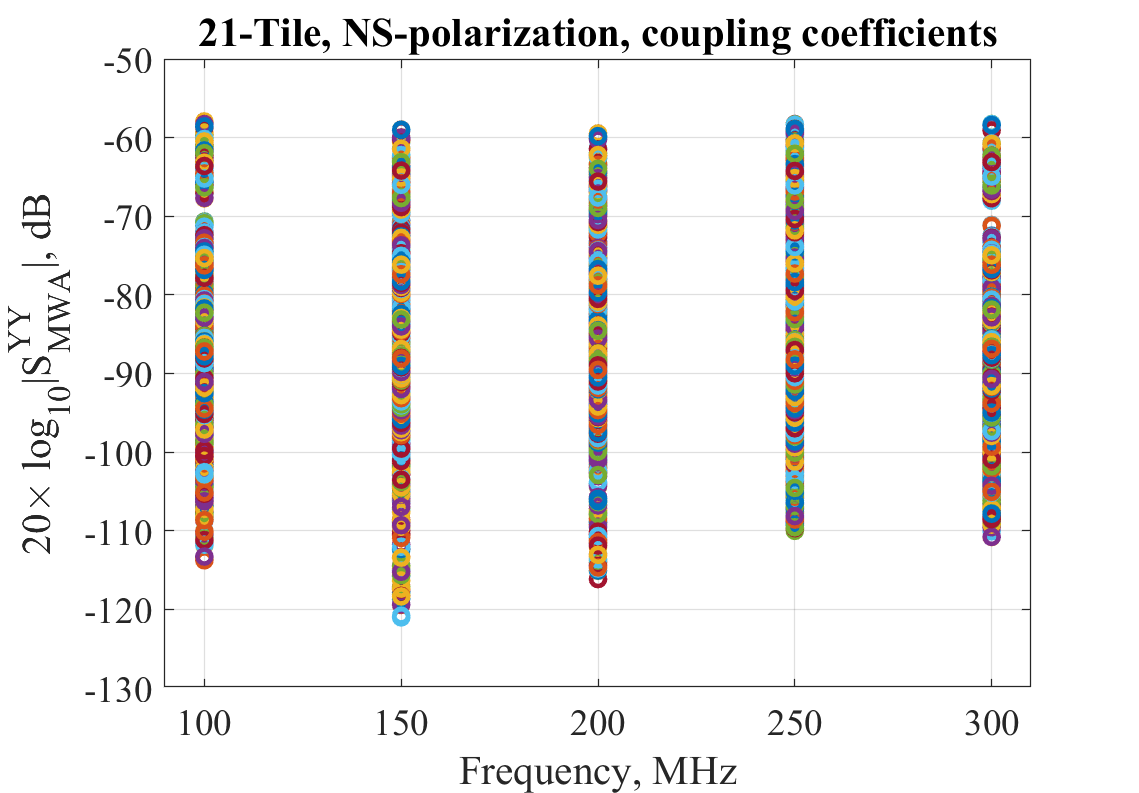}
\caption{21-tile cross-tile coupling coefficients for individual antenna pairs, NS polarization.}
\label{fig:21tile_indivMCs}
\end{figure}

For completeness of provided results, Figure~\ref{fig:21tile_matrix_all} is the full S-parameter matrix for the 21-tile model at 150 MHz showing coupling coefficients between dipoles of all polarizations. Figure~\ref{fig:21tile_matrix_Ypol} highlights the data that represents cross-tile mutual coupling between NS-polarized dipoles. There are several different patterns noticeable in the S-parameters matrix. Due to reciprocity, the matrix is symmetric. Brighter blocks of $16\times16$ dipoles along the diagonal correspond to the coupling within each tile that is stronger because of the close distance between the dipoles in a tile. The square patterns  appear due to the uniform 14~m separation between tiles in the hexagonal array, and the dipoles on the edge of the tiles couple stronger with the neighboring dipoles. In Figure~\ref{fig:21tile_indivMCs}, we show all extracted cross-tile coupling coefficients of all individual dipole pairs (inside the red polygon) at 5 frequency points. In the body of the paper, we show the envelope of this plot in Figure~\ref{fig:21tile_sparams}.

\bsp	
\label{lastpage}
\end{document}